\documentclass[a4paper,fleqn]{latex_styles/cas-sc}
\usepackage[authoryear,longnamesfirst]{natbib}

\usepackage{subfigure}
\usepackage{letltxmacro}
\hypersetup{
  colorlinks=false,
  hidelinks=false,
  pdfborder={0 0 0},
  pdfstartview={FitH}
}
\LetLtxMacro\oldhref\href
\renewcommand{\href}[2]{\oldhref{#1}{\uline{#2}}}
\usepackage{amsmath}
\usepackage{booktabs}
\usepackage{longtable}
\usepackage{array}
\usepackage{multirow}
\usepackage{wrapfig}
\usepackage{colortbl}
\usepackage{pdflscape}
\usepackage{tabu}
\usepackage[normalem]{ulem}
\usepackage{makecell}
\usepackage{xcolor}
\usepackage{caption}
\usepackage{tabularx}
\usepackage{adjustbox}
\usepackage{natbib}
\usepackage{mfirstuc}
\usepackage{float}
\usepackage{placeins}
\usepackage{footnote}
\bibpunct{(}{)}{,}{a}{}{;}

\DeclareCaptionFormat{bold}{\textbf{#1#2}#3}
\setcitestyle{aysep={}}
\setcitestyle{authoryear,open={(},close={)}}

\def\tsc#1{\csdef{#1}{\textsc{\lowercase{#1}}\xspace}}
\tsc{WGM}
\tsc{QE}
\tsc{EP}
\tsc{PMS}
\tsc{BEC}
\tsc{DE}
\begin{document}
% Auto-generated by scripts/get_wid_data.r (do not edit by hand)
\newcommand{\scgYearRange}{1855--2022}

\let\WriteBookmarks\relax
\def\floatpagepagefraction{1}
\def\textpagefraction{.001}

\shorttitle{Capital growth since the 19th century}

\shortauthors{Gordon Getty and Nikita Tkachenko}

\title [mode = title]{Capital growth since the 19th century}

\author[1]{Gordon Getty}[
                        type=editor,
                        role=Researcher,
                        orcid=0000-0002-0939-6932
                        ]

\ead{ggetty@LeakeyFoundation.org}

\author[2]{Nikita Tkachenko}[
                        type=editor,
                        role=Assistant,
                        orcid=0009-0003-8681-3335
                        ]

\cormark[1]

\ead{natkachenko@usfca.edu}

\affiliation[1]{organization={Fellow, UCB},
            addressline={University Avenue and Oxford St},
            city={Berkeley},
            postcode={94720},
            state={California},
            country={United States}}

\affiliation[2]{organization={Graduate Student, University of San Francisco},
            addressline={2130 Fulton St},
            city={San Francisco},
            postcode={94117},
            state={California},
            country={United States}}

\cortext[cor1]{Corresponding author}

\begin{abstract}
We argued in an earlier paper \citep{getty2025sourcescapitalgrowth}, from data reported for net saving and market-value capital by 94 nations over the period 1980-2021, that net saving has had little or no effect on capital growth at the global scale. Here we extend the study backward to the earliest year, in each country, for which those data have been reported or reconstructed.
We find a still smaller margin of possibility (1.6\% versus 6.4\%) of contribution of net saving to capital growth at the collective scale. We also propose and defend an alternative explanation of capital growth, and suggest a larger role for the cash flow concept in economic analysis.
\\
\\
\end{abstract}


\maketitle

\section{Introduction and overview}\label{introduction}

National accounts and modern macroeconomics are founded on the principle that net saving $S_{net}$, within the technological growth rate (warranted rate), tends to give equal growth in market-value capital $K$ before allowance for market noise, which tends to regress to zero over scale and time.\footnote{We follow Keynesian tradition, in this paper, by meaning saving actually invested in new goods and services. More generally, we question Keynes' reasoning that uninvested saving cannot contribute to output, and hence to real saving. It seems to us that saving is invested when appraised present value equals or exceeds investment cost, that such investment opportunities are not always available, and that uninvested saving, in these exceptional but familiar cases, contributes to output by minimizing loss.} We call this common doctrine ``thrift theory''. Then
\begin{equation}
S_{net} = \Delta K, \quad \text{ in thrift theory,}
\label{eq:thrift-saving-capital}
\end{equation}
with the usual understandings that market noise is neglected and that net saving does not exceed the warranted rate.

We first divide by capital (normalize) to avoid nonsense correlation when comparing over time. Define
\begin{equation}
s^* = \frac{S_{net}}{K} \quad \text{and} \quad g(K) = \frac{\Delta K}{K},
\label{eq:normalized-definitions}
\end{equation}
so that Eq.~\eqref{eq:thrift-saving-capital} allows
\begin{equation}
s^* = g(K) \quad \text{and implicitly} \quad \frac{s^*}{g(K)} = 1,\quad \text{ in thrift theory.}
\label{eq:thrift-level}
\end{equation}

We test this prediction against data for $S_{net}$ and $K$ shown in national accounts. Results will show first that net saving, at the scale of all households together, has averaged substantially less than capital growth (57.6\%), and thus cannot have accounted for all of it.\footnote{\citet{picketyCapitalIsBack2014} and \citet{kurz2023market} both tested Eq.~\eqref{eq:thrift-saving-capital} as we do. Both also found \(S_{net}\) significantly less than \(\Delta K\).}

This evidence comparing $s^*$ to $g(K)$ still allows the possibility that a substantial part of overall net saving is recovered in overall capital growth. We test that possibility by recalling that the definitions of covariance, correlation and regression are all founded on the principle that causal relatedness is more clearly revealed in concurrent changes than in concurrent presence. To explore these changes, we note that Eq.~\eqref{eq:thrift-level} implies
\begin{equation}
\Delta s^* = \Delta g(K), \quad \text{giving} \quad \frac{\Delta s^*}{\Delta g(K)} = 1 \quad \text{ in thrift theory.}
\label{eq:thrift-differences}
\end{equation}
For notational convenience, define the ``thrift index'' $\theta$ (theta) by
\begin{equation}
\theta = \frac{\Delta s^*}{\Delta g(K)}, \quad \text{so that Eq.~\eqref{eq:thrift-differences} can be more compactly expressed as }\quad \theta = 1,\quad \text{ in thrift theory.}
\label{eq:thrift-index}
\end{equation}

We have suggested \citep{getty2025sourcescapitalgrowth}, following John Stuart \citet{millPrinciplesPoliticalEconomy1848}, that capital acceleration \(\Delta g(K)\) might also be explained by a concurrent rise in productivity. We called this hypothesis "free growth theory", and tested between the two explanations from data for \(S_{net}\) and \(K\) over the period 1980–2021. Here we extend the test backward to the earliest data we have found for \(S_{net}\) and \(K\).

\section{Test methods and data sources}\label{test-methods}

Our step-by-step methods are shown in our web appendix, \href{web_appendix_scg/}{Capital growth since the 19th century}. Note that we screen results against small denominators, which can magnify fractions and even reverse them in sign. The web appendix shows results when the screen is raised to 0.025, 0.05, 0.075, 0.10, 0.125 and 0.15. Each increase of the screen leaves fewer observations to sample, but raises the informative value of each by showing concurrent changes during progressively steeper positive or negative capital accelerations, when causal relatedness is most clearly revealed.

All data are drawn from the public Piketty-Zucman (PZ) and World Inequality Database (WID) website releases, both of which compile
and collate national account data in currency units of a common year. In some cases, data have been adjusted to fit standards set by the United Nations System of National Accounts (SNA). PZ reports for all years from earliest through 2010. WID covers some of those same years and countries, and continues forward. Data in overlapping years may differ as source countries revise findings. We have preferred the WID data in all these cases.

We measure values from year to year. Thus we measure \(\Delta K_i = K_i - K_{i-1}\) and \(g(K)_i = \frac{\Delta K_i}{K_{i-1}}\) for each successive year \(i\) in each country. \(s^*\) is measured as \(\frac{S_{net,i}}{K_{i-1}}\). We then find \(\frac{s^*_i}{g(K)_i}\), \(\Delta s^*_i = s^*_i - s^*_{i-1}\), \(\Delta g(K)_i = g(K)_i - g(K)_{i-1}\), and \(\theta_i = \frac{\Delta s^*_i}{\Delta g (K)_i}\) for each country and year. For any year and country where data sources did not enable measurement of \(\frac{s^*_i}{g(K)_i}\), that year and country was excluded in measuring \(\frac{s^*}{g(K)}\). Where data sources did not allow measurement of \(\frac{\Delta s^*_i}{\Delta g(K)_i}\), likewise, measurement of \(\theta\) excluded that country for that year.

For the United Kingdom in 1920, the underlying national accounts report alternative versions with and without Northern Ireland. We use the 1920 observation excluding Northern Ireland from first reports in 1855 through 1920, then continue with the series including Northern Ireland from 1921 onward.

\section{Test findings}\label{test-results}

Thrift theory predicts that $\frac{s^*}{g(K)}$, \(\theta\), and the regressions of \(s^*\) on \(g(K)\) and of \(\Delta s^*\) on \(\Delta g(K)\) should all equal unity (the number 1) before allowance for market noise. We test these predictions by deriving yearly and average values of these quantities for each country. Our findings appear in displays below.

\begin{figure}[p]
    \centering
    \includegraphics[width=\textwidth]{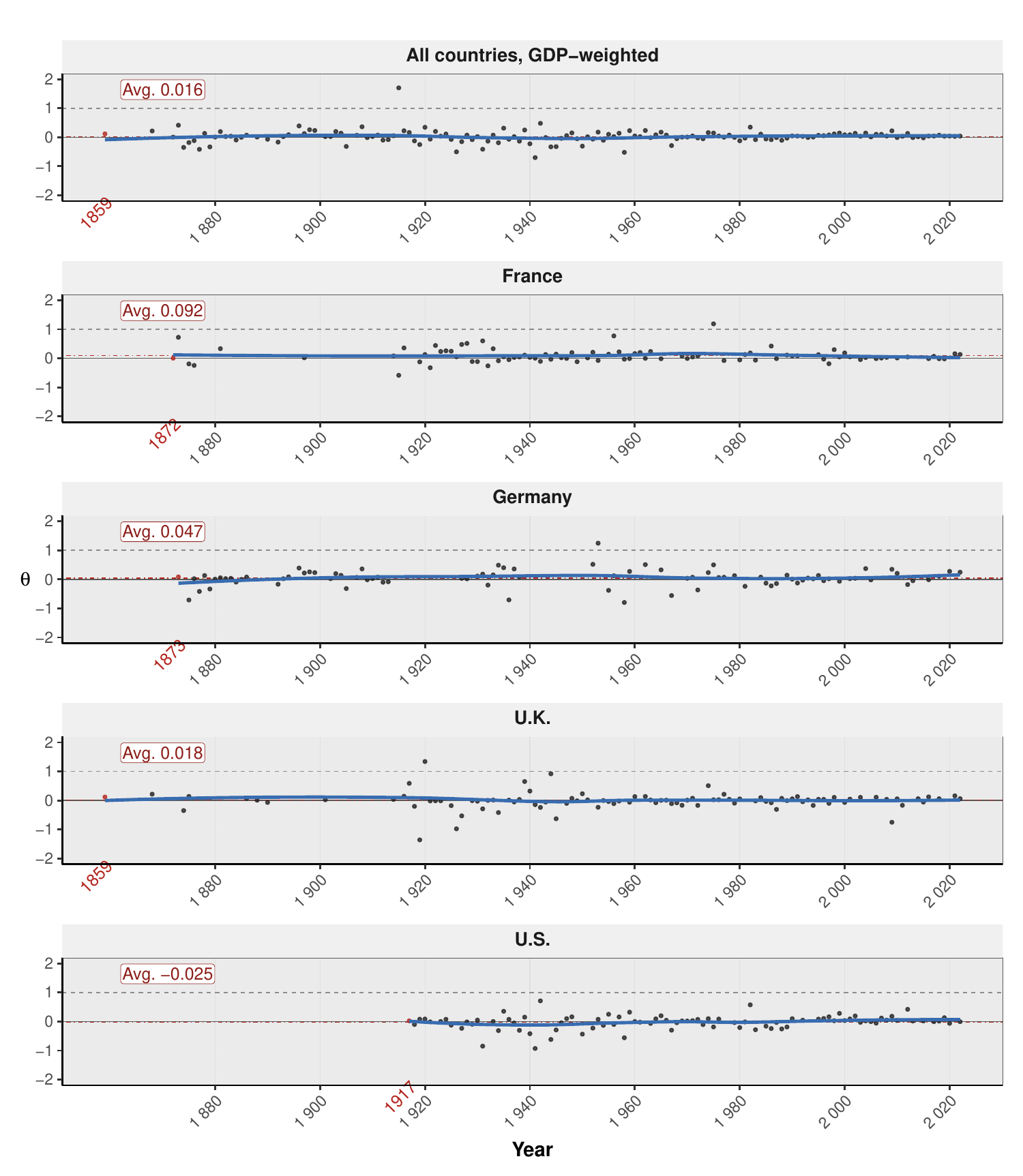}
    \caption{\(\theta\) for all years in countries shown. Screen = 0.01. Dashed line shows the thrift theory prediction \(\theta\) = 1. Blue curves show LOESS best fit; red dot-dash lines show panel averages.}
    \label{fig-country-screen-indicators}
\end{figure}

Table~\ref{tab:interpretive_results} shows GDP-weighted averages of \(s^*/g(K)\) and \(\theta\) across all years and countries as 0.576 and 0.016, respectively. Coefficients of regression of \(s^*\) on \(g(K)\), and of \(\Delta s^*\) on \(\Delta g(K)\), show as 0.141 and 0.018 respectively. Fig.~\ref{fig-country-screen-indicators} plots yearly \(\theta\) for France, Germany, U.K. and U.S., and shows average \(\theta\) over all years at 0.092, 0.047, 0.018 and -0.025 respectively.

We submit that these findings cannot be reconciled with thrift theory.

\section{A parallel argument from consumption restraint}\label{sep12-consumption-restraint}

Under a mild simplifying assumption,\footnote{The share of consumption bought from workers' earnings must hold constant. See Eq.~\eqref{eq:oct4-cash-flow-consumption} below.} the only source of net saving is consumption restraint. \citet{getty2025consumptioncapitalgrowth} followed this logic by defining $c^*$ as $\frac{C}{K}$, and then substituting $-\Delta c^*$ for $\Delta s^*$ in the argument above. We defined $\theta_c = \frac{-\Delta c^*}{\Delta g(K)}$ as the thrift index as found from consumption data.

\begin{figure}[p]
    \centering
    \includegraphics[width=\textwidth]{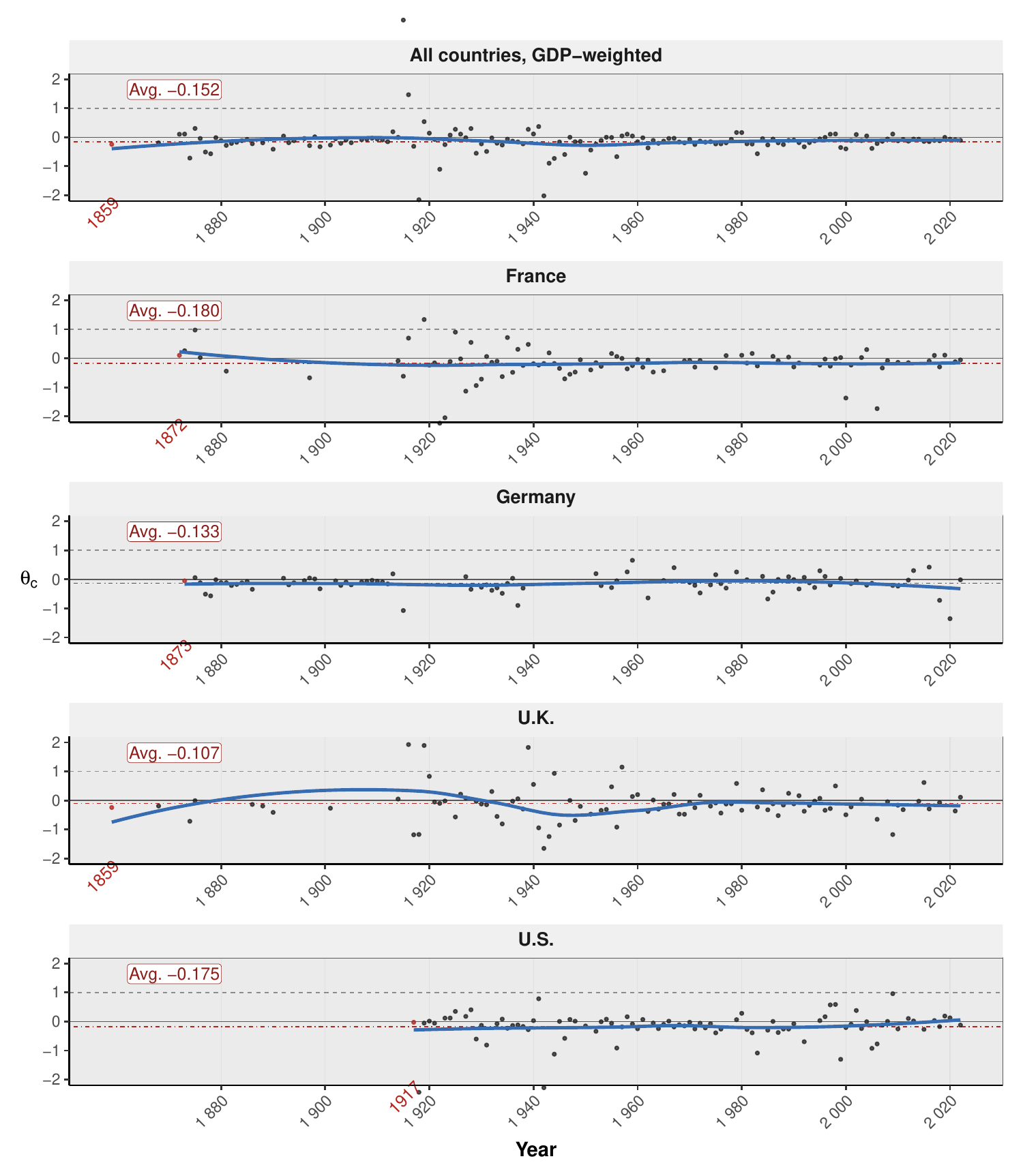}
    \caption{Thrift index (\(\theta_c\)) for all years in countries shown. Screen = 0.01. Dashed line shows the thrift theory prediction \(\theta_c\) = 1. Blue curves show LOESS best fit; red dot-dash lines show panel averages.}
    \label{fig-consumption-indicators}
\end{figure}

Thrift theory, under that simplifying assumption, would predict $\theta_c = 1$. Our paper "Consumption and capital growth" \citep{getty2025consumptioncapitalgrowth} tested that prediction from WID data over the period 1980--2021 to find overall $\theta_c$ at $0.005$. Regressions of $-c^*$ on $g(K)$ and of $-\Delta c^*$ on $\Delta g(K)$ were found at $-0.181$ and $-0.125$ respectively.
Table~\ref{tab:consumption_results} and Fig.~\ref{fig-consumption-indicators} show results of testing from data from WID and PZ combined, again preferring WID when sources disagree.

\section{Free growth theory}\label{free-growth-theory}

In \citet{getty2025sourcescapitalgrowth}, we defined free growth theory as the idea that capital growth is explained by a concurrent rise of productivity in plant and work skills already produced or trained, but did not offer a mathematical description. We will attempt that now.

Cash flow is most familiar to economists as the flow discounted at the time preference rate to find present value, or added to capital growth to find total return.\footnote{All cash flows may be in cash or in kind. "Cash", in this context, means cash value.} It
means outflow of value from an asset to its owner, for consumption or reinvestment elsewhere, less inflow of value into the same asset from outside, as distinct from internal investment in the form of retained output. The usefulness of the concept lies in the reasoning
\begin{equation}
\begin{aligned}
\text{value added} &= \text{value growth} + \text{outflow of value} - \text{inflow of value} \\
&= \text{value growth} + \text{cash flow}.
\label{eq:value-added}
\end{aligned}
\end{equation}
where value added means value created internally.

Value added is also called net output, and value added by physical capital is called net profit. We show net profit as $P$, leaving the qualifier ``net'' implicit, and we show the cash flow drawn from physical capital as $F(K)$.\footnote{We have argued that it is also possible to define the cash flow of human capital \citep{getty2025sourcescapitalgrowth}.} The reasoning of Eq.~\eqref{eq:value-added}, then, as applied to physical capital, finds
\begin{equation}
P = \Delta K + F(K).
\label{eq:total-return}
\end{equation}

Output and cash flow divided by the capital producing them are called rate of return and cash flow rate respectively. If these show as $r(K)$ and $f(K)$, then division of Eq.~\eqref{eq:total-return} by capital $K$ gives
\begin{equation}
r(K) = g(K) + f(K).
\label{eq:return-decomposition}
\end{equation}
For reasons given, we are particularly interested in the relationships among these three variables as they change. We clarify these relationships, as before, by deriving first 
\begin{equation}
\Delta r(K) = \Delta g(K) + \Delta f(K), \quad \text{from which} \quad \frac{\Delta r(K)}{\Delta g(K)} - \frac{\Delta f(K)}{\Delta g(K)} = 1,
\label{eq:return-differences}
\end{equation}
as consequences of Eq.~\eqref{eq:return-decomposition}. We then define
\begin{equation*}
\phi_r = \frac{\Delta r(K)}{\Delta g(K)} \quad \text{and} \quad \theta_r = -\frac{\Delta f(K)}{\Delta g(K)},
\end{equation*}
so that Eq.~\eqref{eq:return-differences} can show more compactly as
\begin{equation}
\phi_r + \theta_r = 1.
\label{eq:return-indices}
\end{equation}
$\phi_r$ and $\theta_r$ respectively can be termed the ``free growth index'' and ``thrift index'' as found from derivation of $r(K)$ in Eq.~\eqref{eq:return-decomposition}.

\(r(K)\), the ratio of the net output of capital to the capital producing it, can be understood as the productivity of capital, and \(\Delta r(K)\) can be understood as a change in that productivity. Thus \(\phi_r\) gives the share of capital acceleration \(\Delta g(K)\) explained by a concurrent change in productivity, while \(\theta_r\) gives the share explained by concurrent restraint in net outflow (cash flow). Let free growth theory mean a hypothesis that capital acceleration $\Delta g(K)$ at the collective scale is explained exclusively by a concurrent productivity gain $\Delta r(K)$. That would give
\begin{equation}
\phi_r = 1, \quad \text{and hence} \quad \theta_r = 0, \quad \text{collectively, \footnotemark\ in free growth theory.}
\label{eq:free-growth-prediction}
\end{equation}
\footnotetext{``Collectively'' here means at the collective or global scale.}
At the same scale, thrift theory would mean the opposite hypothesis
\begin{equation}
\theta_r = 1, \quad \text{and hence} \quad \phi_r = 0, \quad \text{in thrift theory.}
\label{eq:thrift-growth-prediction}
\end{equation}
These hypotheses give extreme cases. We acknowledge that data might suggest something between. 

\section{An approach to testing}\label{approach-to-testing}

We use the term "pay," shown as \(\Pi\), to mean the revenue to human capital in the form of earnings. This flow is also called the labor share of income.

We agree with the consensus view that pay and profit give the only sources
of spending, and that the only dispositions of spending are in investment
and consumption. We express this idea in terms of cash flow by noting
that cash flow means outflow net of all investment, including investment
of pay. As the only remaining disposition is in consumption, we infer
\begin{equation}
\Pi+F(K)=C, \quad \text{collectively},
\label{eq:oct4-cash-flow-consumption}
\end{equation}
where $C$ gives consumption. We specify ``collectively'' to recognize
the possibility that payments at smaller scales might also be passed
among households in transfer payments.

Our paper "A separate way to measure rate of return"\footnote{\citet{getty2025separatewaymeasurerateofreturn}.} found the same conclusion. Here we add a refinement omitted in that paper. Eq.~\eqref{eq:oct4-cash-flow-consumption} reasons from sources and dispositions of payment. It is reasonable to acknowledge unpaid consumption services (UCS) as in parenting or other care of self and family. UCS could thus be recognized as an imputed component of both pay $\Pi$ and consumption $C$. Eq.~\eqref{eq:oct4-cash-flow-consumption} might be understood with or without this imputation.

% Source PDF page 4 (yellow handwritten page).
% Circled instruction: "insert pt 1 on p 4".
% Margin instruction: "add".
\section{A closer argument for Eq.~\eqref{eq:oct4-cash-flow-consumption}}
\label{oct17-closer-argument}

We will rely on Eq.~\eqref{eq:oct4-cash-flow-consumption}, which we call the cash flow rule, and will now defend it more rigorously.
\citet{schultzInvestmentHumanCapital1961} distinguished between consumption invested in human capital in such forms as teaching, nurture and parenting, and ``pure'' consumption exhausted from the economy in satisfying tastes. We notate these flows as $C_s$ and $C_p$ respectively, where the subscript $s$ means ``saved''. Thus
\begin{equation}
C_s + C_p = C\ .
\label{eq:oct17-14}
\end{equation}

It is reasonable and customary to treat consumption, and implicitly $C_s$, as the only medium of investment in human capital from outside, as distinct from internal investment, which we call self-invested work, in such forms as learning to complement teaching (\citet{schultzInvestmentHumanCapital1961}).
% PROOFREAD: Final faint parenthesis after "thinking" is not securely legible.
% Tentative reading: "including oneself"; please check against PDF page 4.
% Margin insertion "such forms" has been incorporated in the preceding sentence.

% Source PDF page 5 (yellow handwritten page).
% Circled instruction: "insert pt 2 on p 4".
Cash flow $F(K)$, meanwhile, is defined as net outflow of capital value $K$ from assets to owners. If outflow and inflow show as $F(K)_+$ and $F(K)_-$ respectively, we may write
% PROOFREAD: "inflow of capital into K" is the best visual reading of a very faint line; check original. It appears inconsistent with the following equation.
\begin{equation}
F(K) = F(K)_+ - F(K)_-.
\label{eq:oct17-15}
\end{equation}

In general, outflow $F_+$ drawn from a particular asset might be reinvested in another asset, or invested in human capital through invested consumption $C_s$ in payment for textbooks and tuition, say, or might be exhausted from the economy in pure consumption $C_p$, or passed from one household to another in transfer payments. To limit these possibilities, take all physical capital $K$ in the economy as a single collective set, so that reinvestment in other assets is foreclosed, and consider all households together, where transfers given and received offset to zero. The possible remaining dispositions of $F(K)_+$ are in invested consumption $C_s$ and pure consumption $C_p$. $F(K)_+$ does
not provide the only source of inflow into $C_s$ and $C_p$, however, as pay $\Pi$ provides another. Let the shares sourced from $F(K)_+$ show as $C_s(F(K))$ and $C_p(F(K))$ respectively. Then
\begin{equation}
F(K)_+ = C_s(F(K)) + C_p(F(K)), \quad \text{collectively}.
\label{eq:oct17-16}
\end{equation}
% Small handwritten spacing instruction before "collectively": "3 space".

By Eq.~\eqref{eq:oct17-14}, the right side of this equation gives the entirety of consumption sourced from cash flow \(F(K)\). If this shows as $C(F(K))$, Eq.~\eqref{eq:oct17-16} may be expressed more compactly as
% PROOFREAD: "sourced from pay" is retained as written, despite the apparent inconsistency with C(F(K)).
\begin{equation}
F(K)_+ = C(F(K)), \quad \text{collectively}.
\label{eq:oct17-17}
\end{equation}
% Small handwritten spacing instruction before "collectively": "3 space".

The only possible dispositions of pay $\Pi$, meanwhile, at the collective scale where transfer is ruled out, are in investment in physical capital in the form of inflow $F(K)_-$, investment in human capital in the form of invested consumption $C_s$, and in exhaustion of value in taste satisfaction through pure consumption $C_p$. By
analogy with Eq.~\eqref{eq:oct17-16}, let the shares of
these sourced from pay show respectively
as $F(K)_{-}(\Pi)$, $C_s(\Pi)$ and $C_p(\Pi)$. Then
\begin{equation}
\Pi = F(K)_{-}(\Pi) + C_s(\Pi) + C_p(\Pi), \quad \text{collectively}.
\label{eq:oct17-18}
\end{equation}
As with Eq.~\eqref{eq:oct17-17}, we may reason that
$C_s(\Pi)+C_p(\Pi)$ gives the whole of consumption
sourced from pay, and may write
this sum as $C(\Pi)$. Thus Eq.~\eqref{eq:oct17-18} may
be more simply put as
\begin{equation}
\Pi = F(K)_{-}(\Pi) + C(\Pi), \quad \text{collectively}.
\label{eq:oct17-19}
\end{equation}
By the understanding that $K$ here means
the set of all physical capital in the
economy, again, outflow $F(K)_{+}$ is not an
available source for inflow or capital
investment $F(K)_{-}$. The only remaining
source is pay $\Pi$. Then
\begin{equation}
F(K)_{-} = F(K)_{-}(\Pi), \quad \text{collectively}.
\label{eq:oct17-20}
\end{equation}
%
% ======================================================================
% PDF page 8: yellow handwritten page.
% Circled instruction: ``insert pt 5 in p 4''.
% Source arrows request ``3 spaces'' before ``collectively''.
% ======================================================================
By Eqs.~\eqref{eq:oct17-15}, \eqref{eq:oct17-17} and \eqref{eq:oct17-20}, now, we find
\begin{equation}
\begin{aligned}
F(K) &= F(K)_{+} - F(K)_{-} \\
     &= C(F(K)) - F(K)_{-}(\Pi), \quad \text{collectively}.
\end{aligned}
\label{eq:oct17-21}
\end{equation}
This and Eq.~\eqref{eq:oct17-19} together give
\begin{equation}
\begin{aligned}
F(K)+\Pi &= C(F(K))-F(K)_{-}(\Pi)+F(K)_{-}(\Pi)+C(\Pi) \\
         &= C(F(K))+C(\Pi), \quad \text{collectively}.
\end{aligned}
\label{eq:oct17-22}
\end{equation}
As $F(K)$ and $\Pi$ give the only sources
of consumption, finally, we get
\begin{equation}
C(F(K))+C(\Pi)=C, \quad \text{collectively}.
\label{eq:oct17-23}
\end{equation}
% Proofreading: the source says ``Eq. 20'' here. This appears inconsistent
% with the displayed derivation; the handwritten reference is preserved.
Substitution of this in Eq.~\eqref{eq:oct17-22} arrives at
\begin{equation*}
F(K)+\Pi=C, \quad \text{collectively},
\end{equation*}
confirming Eq.~\eqref{eq:oct4-cash-flow-consumption}.

\section{Test methods}\label{free-growth-test-methods}

National accounts in many nations report compensation of workers and mixed income.\footnote{As with income of partnerships and proprietorships.} WID constructs labor income, which we show as pay $\Pi$, as compensation of workers plus 70\% of net mixed income.

PZ also reports total labor income, including an estimated labor component of self-employment income, but its estimation method varies by country and source rather than applying WID's fixed 70\% allocation. We therefore do not merge the PZ and WID pay series. Doing so would introduce a methodological break that could create artificial year-to-year changes in $F(K)$ and the indices derived from it. Among the four countries displayed, the merger would extend only Germany's plotted range, from 1971 back to 1952, adding ten observations that pass our 0.01 screen. Among the other countries for which the PZ workbooks provide a comparable total-labor-income series, only Italy has earlier raw coverage in PZ than in WID: PZ begins in 1960 and WID in 1970. These comparisons concern the raw pay series; the number of usable observations is further limited by the availability of the other variables and of consecutive years, and by the screen.

Thus we use data for consumption $C$ from WID where reported and from PZ for countries and years not covered in WID, while using WID alone for pay $\Pi$. We find collective $F(K)$ by Eq.~\eqref{eq:oct4-cash-flow-consumption} for these countries and years. We then apply values already found for collective $K$, $\Delta K$, $g(K)$ and $\Delta g(K)$ to find collective $P$ in Eq.~\eqref{eq:total-return}, $f(K)$ and $r(K)$ in Eq.~\eqref{eq:return-decomposition}, $\Delta f(K)$ and $\Delta r(K)$ in Eq.~\eqref{eq:return-differences}, then $\phi_r$ and $\theta_r$ in Eq.~\eqref{eq:return-indices}.

\FloatBarrier

\clearpage
\begingroup
\fontsize{9}{9.8}\selectfont
\renewcommand{\arraystretch}{1.04}
\setlength{\tabcolsep}{1.5pt}
\setlength{\LTleft}{0pt}
\setlength{\LTright}{0pt}
\setlength{\LTcapwidth}{\linewidth}
\begin{longtable}{lrrccccccccc}
\caption{$g(K)$, $s^*/g(K)$, $\theta$ and associated coefficients of regression averaged over all years reported for each country. Screen = 0.01. $H_0$ per thrift theory: $s^*/g(K), \theta$ and each coefficient of regression $\cong 1$.}\label{tab:interpretive_results}\\
\toprule
 & & & & \multicolumn{4}{c}{$s^*$ on $g(K)$} & \multicolumn{4}{c}{$\Delta s^*$ on $\Delta g(K)$}\\
\cmidrule(lr){5-8} \cmidrule(lr){9-12}
Country & $g(K)$ & $s^*/g(K)$ & $\theta$ & Coef. & S.E. & Obs. & R$^2$ & Coef. & S.E. & Obs. & R$^2$\\
\midrule
\endfirsthead
\caption*{\textbf{\tablename~\thetable}\ (cont.)}\\
\toprule
 & & & & \multicolumn{4}{c}{$s^*$ on $g(K)$} & \multicolumn{4}{c}{$\Delta s^*$ on $\Delta g(K)$}\\
\cmidrule(lr){5-8} \cmidrule(lr){9-12}
Country & $g(K)$ & $s^*/g(K)$ & $\theta$ & Coef. & S.E. & Obs. & R$^2$ & Coef. & S.E. & Obs. & R$^2$\\
\midrule
\endhead
Armenia & 0.027 & -0.035 & -0.018 & 0.008 & (0.053) & 29 & 0.001 & -0.005 & (0.023) & 31 & 0.002\\
Aruba & 0.024 & 0.190 & 0.062 & 0.029$^{**}$ & (0.010) & 25 & 0.161 & 0.022$^{***}$ & (0.004) & 23 & 0.375\\
Australia & 0.049 & 0.425 & 0.053 & 0.066 & (0.044) & 54 & 0.029 & 0.023$^{*}$ & (0.011) & 50 & 0.099\\
Austria & 0.027 & 0.752 & 0.051 & 0.051$^{**}$ & (0.024) & 23 & 0.167 & 0.041$^{*}$ & (0.020) & 21 & 0.218\\
Azerbaijan & 0.052 & 0.402 & 0.117 & 0.220$^{***}$ & (0.043) & 30 & 0.366 & 0.025 & (0.042) & 28 & 0.019\\
Bahrain & 0.034 & 0.190 & 0.068 & 0.087$^{***}$ & (0.022) & 12 & 0.499 & 0.096$^{**}$ & (0.031) & 13 & 0.724\\
Belarus & 0.019 & 0.195 & 0.069 & 0.013 & (0.031) & 30 & 0.012 & 0.009 & (0.018) & 29 & 0.014\\
Belgium & 0.031 & 0.781 & 0.061 & 0.164$^{***}$ & (0.044) & 29 & 0.197 & 0.064$^{***}$ & (0.019) & 29 & 0.308\\
Bolivia & 0.054 & 0.288 & -0.028 & 0.130$^{***}$ & (0.036) & 23 & 0.240 & 0.015 & (0.029) & 23 & 0.014\\
Botswana & 0.048 & 0.672 & 0.061 & 0.118$^{***}$ & (0.029) & 31 & 0.297 & 0.072$^{***}$ & (0.018) & 30 & 0.331\\
Brazil & 0.025 & 0.276 & 0.115 & 0.160$^{***}$ & (0.033) & 20 & 0.630 & 0.091$^{***}$ & (0.024) & 18 & 0.499\\
British Virgin Islands & 0.124 & 0.430 & 0.045 & 0.029$^{***}$ & (0.006) & 25 & 0.239 & 0.018 & (0.011) & 23 & 0.378\\
Bulgaria & 0.065 & 0.396 & 0.125 & 0.127$^{***}$ & (0.016) & 23 & 0.351 & 0.111 & (0.070) & 19 & 0.358\\
Burkina Faso & 0.055 & 0.318 & 0.015 & -0.067 & (0.045) & 21 & 0.083 & 0.023 & (0.016) & 21 & 0.073\\
Cameroon & 0.045 & 0.509 & 0.015 & 0.032 & (0.044) & 24 & 0.039 & 0.055$^{**}$ & (0.020) & 26 & 0.321\\
Canada & 0.049 & 0.602 & 0.107 & 0.069 & (0.056) & 53 & 0.019 & 0.012 & (0.022) & 43 & 0.007\\
Cape Verde & 0.027 & 0.225 & 0.028 & 0.018$^{*}$ & (0.010) & 13 & 0.053 & 0.015$^{***}$ & (0.004) & 11 & 0.102\\
Chile & 0.037 & 0.465 & 0.014 & 0.091 & (0.096) & 23 & 0.093 & 0.024 & (0.020) & 19 & 0.047\\
China & 0.103 & 0.681 & 0.086 & 0.117$^{**}$ & (0.052) & 30 & 0.135 & 0.054$^{***}$ & (0.013) & 23 & 0.359\\
Colombia & 0.047 & 0.280 & 0.050 & 0.043 & (0.061) & 27 & 0.047 & 0.020 & (0.022) & 24 & 0.057\\
Costa Rica & 0.025 & 0.562 & 0.005 & -0.000 & (0.007) & 8 & 0.000 & 0.014 & (0.014) & 7 & 0.220\\
Croatia & 0.034 & 0.035 & 0.016 & 0.047$^{**}$ & (0.017) & 23 & 0.113 & 0.009 & (0.018) & 21 & 0.015\\
Curaçao & 0.010 & 0.535 & 0.044 & 0.056 & (0.058) & 20 & 0.033 & 0.014 & (0.031) & 20 & 0.020\\
Cyprus & 0.037 & 0.136 & 0.113 & 0.039 & (0.033) & 22 & 0.053 & 0.007 & (0.044) & 24 & 0.003\\
Czechia & 0.008 & 0.165 & 0.022 & 0.028$^{*}$ & (0.014) & 26 & 0.085 & 0.048$^{**}$ & (0.018) & 22 & 0.241\\
Côte d’Ivoire & 0.065 & 0.441 & 0.065 & 0.154$^{**}$ & (0.063) & 32 & 0.202 & 0.022 & (0.027) & 27 & 0.021\\
Denmark & 0.035 & 0.335 & 0.037 & 0.050$^{**}$ & (0.019) & 40 & 0.238 & 0.024$^{**}$ & (0.010) & 38 & 0.160\\
Dominican Republic & 0.057 & 0.307 & 0.055 & 0.048$^{*}$ & (0.027) & 28 & 0.110 & 0.038$^{***}$ & (0.011) & 27 & 0.226\\
Ecuador & 0.053 & 0.244 & 0.058 & 0.053$^{***}$ & (0.011) & 15 & 0.511 & 0.035$^{***}$ & (0.004) & 10 & 0.721\\
Egypt & 0.060 & 0.406 & 0.061 & 0.051 & (0.033) & 22 & 0.044 & 0.021$^{***}$ & (0.007) & 21 & 0.156\\
El Salvador & 0.033 & -0.332 & -0.445 & -0.090 & (0.183) & 7 & 0.058 & -0.028 & (0.025) & 6 & 0.049\\
Estonia & 0.039 & 0.471 & -0.058 & 0.133$^{***}$ & (0.023) & 27 & 0.570 & 0.061$^{*}$ & (0.033) & 23 & 0.151\\
Finland & 0.038 & 0.430 & -0.008 & 0.149$^{**}$ & (0.059) & 36 & 0.281 & 0.022 & (0.018) & 29 & 0.047\\
France & 0.022 & 0.565 & 0.092 & 0.173$^{***}$ & (0.031) & 124 & 0.302 & 0.035$^{**}$ & (0.017) & 88 & 0.091\\
Germany & 0.040 & 0.617 & 0.047 & 0.296$^{***}$ & (0.072) & 119 & 0.329 & 0.056$^{***}$ & (0.009) & 86 & 0.269\\
Greece & 0.018 & 0.127 & 0.129 & 0.134$^{***}$ & (0.026) & 24 & 0.541 & 0.070$^{***}$ & (0.020) & 24 & 0.438\\
Guatemala & 0.037 & -0.565 & 0.026 & -0.036 & (0.084) & 19 & 0.029 & 0.004 & (0.040) & 16 & 0.002\\
Guinea & 0.196 & 0.670 & 0.077 & 0.032$^{***}$ & (0.010) & 19 & 0.142 & 0.042$^{***}$ & (0.009) & 16 & 0.327\\
Honduras & 0.034 & -0.093 & 0.107 & 0.019 & (0.054) & 16 & 0.008 & 0.020 & (0.020) & 15 & 0.041\\
Hong Kong & 0.061 & 0.440 & 0.078 & 0.191$^{***}$ & (0.045) & 41 & 0.307 & 0.051$^{**}$ & (0.021) & 37 & 0.199\\
Hungary & 0.033 & 0.274 & 0.107 & 0.196$^{***}$ & (0.039) & 23 & 0.447 & 0.069$^{**}$ & (0.027) & 20 & 0.288\\
Iceland & 0.037 & 0.214 & 0.062 & 0.138$^{***}$ & (0.032) & 19 & 0.510 & 0.084$^{***}$ & (0.022) & 17 & 0.501\\
India & 0.074 & 0.516 & 0.188 & 0.087 & (0.075) & 23 & 0.176 & 0.025$^{*}$ & (0.012) & 15 & 0.118\\
Indonesia & 0.062 & 0.058 & 0.006 & 0.021 & (0.026) & 5 & 0.068 & 0.003 & (0.019) & 5 & 0.011\\
Iran & 0.029 & 0.529 & 0.043 & 0.188$^{***}$ & (0.048) & 20 & 0.371 & 0.097$^{***}$ & (0.033) & 24 & 0.308\\
Ireland & 0.057 & 0.525 & 0.066 & 0.112$^{***}$ & (0.021) & 25 & 0.537 & 0.073$^{***}$ & (0.022) & 22 & 0.395\\
Israel & 0.054 & 0.367 & 0.036 & -0.025 & (0.055) & 25 & 0.010 & 0.034 & (0.021) & 21 & 0.145\\
Italy & 0.025 & 0.140 & 0.049 & 0.081$^{***}$ & (0.028) & 35 & 0.177 & 0.015 & (0.014) & 19 & 0.073\\
Japan & 0.025 & 0.208 & 0.044 & 0.116$^{***}$ & (0.026) & 28 & 0.339 & 0.013 & (0.011) & 26 & 0.049\\
Kazakhstan & 0.022 & 0.446 & 0.022 & 0.153$^{***}$ & (0.055) & 29 & 0.306 & 0.046$^{**}$ & (0.019) & 28 & 0.142\\
Kuwait & 0.044 & 1.335 & 0.241 & 0.174$^{*}$ & (0.082) & 18 & 0.199 & 0.048$^{*}$ & (0.026) & 17 & 0.064\\
Kyrgyzstan & 0.024 & -0.207 & -0.075 & -0.043 & (0.033) & 31 & 0.038 & -0.019 & (0.020) & 30 & 0.025\\
Latvia & 0.044 & -0.013 & -0.003 & 0.032 & (0.113) & 26 & 0.005 & -0.056$^{***}$ & (0.020) & 26 & 0.106\\
Lithuania & 0.048 & 0.217 & 0.003 & -0.048 & (0.080) & 25 & 0.028 & 0.019 & (0.045) & 20 & 0.018\\
Luxembourg & 0.047 & 0.621 & 0.019 & 0.010 & (0.035) & 24 & 0.001 & -0.021 & (0.026) & 22 & 0.044\\
Malaysia & 0.044 & 1.047 & 0.233 & 0.232$^{***}$ & (0.074) & 16 & 0.237 & 0.063 & (0.052) & 11 & 0.204\\
Malta & 0.043 & 0.622 & 0.061 & 0.420$^{***}$ & (0.125) & 24 & 0.384 & 0.103 & (0.064) & 22 & 0.098\\
Mauritius & 0.029 & 0.005 & -0.006 & -0.007 & (0.005) & 8 & 0.231 & -0.007$^{***}$ & (0.002) & 8 & 0.613\\
Mexico & 0.029 & 0.284 & 0.086 & 0.063$^{*}$ & (0.033) & 25 & 0.132 & 0.042 & (0.030) & 21 & 0.148\\
Moldova & -0.004 & -0.275 & -0.037 & -0.043 & (0.067) & 31 & 0.042 & 0.035 & (0.021) & 32 & 0.174\\
Mongolia & 0.055 & 0.273 & 0.180 & 0.155$^{***}$ & (0.033) & 15 & 0.723 & 0.108$^{***}$ & (0.015) & 15 & 0.674\\
Morocco & 0.054 & 0.625 & 0.013 & 0.057$^{**}$ & (0.022) & 22 & 0.132 & 0.032$^{***}$ & (0.011) & 18 & 0.251\\
Netherlands & 0.026 & 0.523 & -0.012 & 0.061$^{*}$ & (0.031) & 38 & 0.155 & 0.023 & (0.018) & 28 & 0.036\\
New Zealand & 0.044 & 0.566 & -0.043 & 0.011 & (0.031) & 32 & 0.003 & 0.044 & (0.027) & 33 & 0.051\\
Nicaragua & 0.035 & 0.144 & 0.138 & 0.036 & (0.053) & 14 & 0.041 & 0.023 & (0.018) & 14 & 0.086\\
Niger & 0.067 & 0.254 & 0.017 & 0.032 & (0.030) & 27 & 0.032 & 0.036$^{***}$ & (0.010) & 23 & 0.220\\
Norway & 0.046 & 0.899 & -0.054 & -0.001 & (0.043) & 42 & 0.000 & -0.045$^{**}$ & (0.022) & 38 & 0.086\\
Peru & 0.048 & 0.462 & 0.063 & 0.031 & (0.038) & 15 & 0.086 & 0.022 & (0.021) & 12 & 0.133\\
Philippines & 0.050 & 0.999 & 0.010 & 0.121 & (0.093) & 28 & 0.106 & 0.041 & (0.035) & 22 & 0.076\\
Poland & 0.063 & 0.344 & 0.105 & 0.050 & (0.036) & 30 & 0.068 & 0.066$^{*}$ & (0.033) & 21 & 0.371\\
Portugal & 0.030 & 0.027 & 0.029 & 0.060$^{**}$ & (0.026) & 22 & 0.203 & 0.010 & (0.015) & 19 & 0.023\\
Qatar & 0.085 & -0.313 & 0.205 & 0.294$^{***}$ & (0.041) & 20 & 0.756 & 0.129$^{*}$ & (0.073) & 19 & 0.279\\
Romania & 0.011 & -0.118 & 0.012 & 0.007 & (0.028) & 30 & 0.002 & -0.007 & (0.011) & 27 & 0.018\\
Russia & 0.045 & 0.308 & 0.111 & 0.048$^{*}$ & (0.026) & 22 & 0.123 & 0.027 & (0.024) & 24 & 0.034\\
Saudi Arabia & 0.068 & 0.615 & 0.117 & 0.193$^{***}$ & (0.051) & 17 & 0.373 & 0.092$^{**}$ & (0.037) & 14 & 0.330\\
Senegal & 0.051 & 0.119 & -0.084 & -0.153 & (0.091) & 8 & 0.320 & -0.058$^{*}$ & (0.022) & 5 & 0.396\\
Serbia & 0.052 & -0.114 & -0.072 & -0.137 & (0.086) & 25 & 0.135 & -0.090 & (0.076) & 20 & 0.215\\
Slovakia & 0.028 & 0.153 & 0.051 & 0.148$^{***}$ & (0.050) & 24 & 0.194 & 0.058 & (0.046) & 26 & 0.109\\
Slovenia & 0.033 & 0.443 & 0.101 & 0.166$^{***}$ & (0.035) & 26 & 0.544 & 0.099$^{**}$ & (0.041) & 18 & 0.399\\
South Africa & 0.024 & 0.331 & 0.026 & 0.063 & (0.037) & 18 & 0.186 & 0.039$^{**}$ & (0.017) & 23 & 0.178\\
South Korea & 0.080 & 0.541 & -0.003 & 0.081$^{**}$ & (0.034) & 51 & 0.088 & 0.059$^{***}$ & (0.017) & 41 & 0.390\\
Spain & 0.032 & 0.322 & 0.067 & 0.084$^{***}$ & (0.013) & 22 & 0.563 & 0.028$^{**}$ & (0.012) & 21 & 0.187\\
Sweden & 0.041 & 0.711 & 0.045 & 0.096$^{**}$ & (0.037) & 64 & 0.104 & 0.053$^{***}$ & (0.018) & 56 & 0.162\\
Switzerland & 0.032 & 0.420 & 0.093 & 0.038 & (0.032) & 24 & 0.027 & 0.075$^{*}$ & (0.040) & 22 & 0.149\\
Trinidad \& Tobago & 0.031 & 0.008 & 0.002 & 0.005$^{***}$ & (0.001) & 22 & 0.672 & 0.002$^{***}$ & (0.000) & 20 & 0.624\\
Tunisia & 0.041 & -0.013 & 0.064 & 0.086 & (0.056) & 27 & 0.090 & 0.004 & (0.018) & 25 & 0.002\\
Turkey & 0.068 & 1.148 & 0.022 & -0.045 & (0.153) & 13 & 0.014 & -0.042 & (0.027) & 11 & 0.137\\
USA & 0.037 & 0.553 & -0.025 & 0.032 & (0.027) & 142 & 0.017 & 0.004 & (0.010) & 89 & 0.002\\
Ukraine & -0.017 & 0.081 & 0.092 & 0.012 & (0.022) & 30 & 0.009 & 0.020$^{*}$ & (0.010) & 28 & 0.115\\
United Kingdom & 0.021 & 0.429 & 0.018 & 0.096$^{***}$ & (0.023) & 142 & 0.165 & 0.008 & (0.014) & 97 & 0.006\\
Uruguay & 0.005 & 0.630 & 0.060 & 0.022 & (0.018) & 6 & 0.195 & 0.026 & (0.014) & 5 & 0.448\\
Uzbekistan & 0.052 & 0.122 & 0.128 & 0.163 & (0.123) & 12 & 0.124 & 0.018 & (0.032) & 9 & 0.060\\
Vanuatu & 0.048 & 0.280 & 0.034 & 0.060 & (0.058) & 19 & 0.113 & 0.028$^{**}$ & (0.012) & 15 & 0.258\\
Venezuela & -0.023 & 0.196 & 0.125 & 0.137$^{***}$ & (0.025) & 22 & 0.542 & 0.037 & (0.030) & 22 & 0.126\\
ALL COUNTRIES\textsuperscript{a} & 0.033 & 0.576 & 0.016 & 0.141$^{***}$ & (0.015) & 2744 & 0.175 & 0.018$^{***}$ & (0.007) & 2347 & 0.025\\
\bottomrule
\end{longtable}
\begin{flushleft}
\footnotesize \textsuperscript{a} GDP-weighted.\\
\footnotesize \emph{Note:} Three, two and one asterisks denote $p<0.01$, $p<0.05$ and $p<0.10$, respectively.
\end{flushleft}
\endgroup

\clearpage
\begingroup
\fontsize{9}{9.8}\selectfont
\renewcommand{\arraystretch}{1.04}
\setlength{\tabcolsep}{1.5pt}
\setlength{\LTleft}{0pt}
\setlength{\LTright}{0pt}
\setlength{\LTcapwidth}{\linewidth}
\begin{longtable}{lrrcccccccc}
\caption{$g(K)$, $\theta_c$ and associated coefficients of regression averaged over all years reported for each country. Screen = 0.01. $H_0$ per thrift theory: $\theta_c$ and each coefficient of regression $\cong 1$.}\label{tab:consumption_results}\\
\toprule
 & & & \multicolumn{4}{c}{$-c^*$ on $g(K)$} & \multicolumn{4}{c}{$-\Delta c^*$ on $\Delta g(K)$}\\
\cmidrule(lr){4-7} \cmidrule(lr){8-11}
Country & $g(K)$ & $\theta_c$ & Coef. & S.E. & Obs. & R$^2$ & Coef. & S.E. & Obs. & R$^2$\\
\midrule
\endfirsthead
\caption*{\textbf{\tablename~\thetable}\ (cont.)}\\
\toprule
 & & & \multicolumn{4}{c}{$-c^*$ on $g(K)$} & \multicolumn{4}{c}{$-\Delta c^*$ on $\Delta g(K)$}\\
\cmidrule(lr){4-7} \cmidrule(lr){8-11}
Country & $g(K)$ & $\theta_c$ & Coef. & S.E. & Obs. & R$^2$ & Coef. & S.E. & Obs. & R$^2$\\
\midrule
\endhead
Armenia & 0.027 & -0.163 & -0.202$^{**}$ & (0.076) & 29 & 0.363 & -0.151$^{***}$ & (0.023) & 31 & 0.691\\
Aruba & 0.024 & -0.163 & -0.186$^{**}$ & (0.075) & 25 & 0.288 & -0.166$^{***}$ & (0.022) & 23 & 0.817\\
Australia & 0.049 & -0.073 & -0.125 & (0.131) & 54 & 0.019 & -0.078$^{***}$ & (0.011) & 50 & 0.450\\
Austria & 0.027 & -0.109 & -0.150 & (0.164) & 23 & 0.059 & -0.115$^{***}$ & (0.015) & 21 & 0.674\\
Azerbaijan & 0.052 & -0.066 & 0.025 & (0.026) & 30 & 0.006 & -0.133$^{***}$ & (0.031) & 28 & 0.450\\
Bahrain & 0.034 & -0.071 & -0.077$^{*}$ & (0.038) & 12 & 0.345 & -0.079$^{***}$ & (0.016) & 13 & 0.690\\
Belarus & 0.019 & -0.114 & -0.139$^{***}$ & (0.023) & 30 & 0.499 & -0.125$^{***}$ & (0.011) & 29 & 0.776\\
Belgium & 0.031 & -0.070 & -0.309 & (0.183) & 29 & 0.064 & -0.052$^{*}$ & (0.027) & 29 & 0.137\\
Bolivia & 0.054 & -0.112 & -0.050 & (0.191) & 23 & 0.003 & -0.177$^{***}$ & (0.055) & 23 & 0.349\\
Botswana & 0.048 & -0.077 & -0.207$^{***}$ & (0.051) & 31 & 0.226 & -0.160$^{***}$ & (0.037) & 30 & 0.522\\
Brazil & 0.025 & -0.130 & -0.242$^{**}$ & (0.109) & 20 & 0.198 & -0.132$^{**}$ & (0.047) & 18 & 0.423\\
British Virgin Islands & 0.124 & -0.064 & -0.066$^{***}$ & (0.014) & 25 & 0.161 & -0.043 & (0.025) & 23 & 0.414\\
Bulgaria & 0.065 & -0.297 & -0.265 & (0.338) & 23 & 0.077 & -0.218 & (0.133) & 19 & 0.460\\
Burkina Faso & 0.055 & -0.076 & -0.288$^{***}$ & (0.088) & 21 & 0.308 & -0.122$^{***}$ & (0.024) & 21 & 0.522\\
Cameroon & 0.045 & -0.297 & -0.279$^{**}$ & (0.117) & 24 & 0.205 & -0.211$^{***}$ & (0.054) & 26 & 0.548\\
Canada & 0.049 & -0.086 & -0.099 & (0.267) & 53 & 0.002 & -0.164$^{***}$ & (0.032) & 43 & 0.382\\
Cape Verde & 0.027 & -0.251 & -0.238$^{***}$ & (0.033) & 13 & 0.879 & -0.244$^{***}$ & (0.008) & 11 & 0.986\\
Chile & 0.037 & -0.272 & -0.331$^{**}$ & (0.130) & 23 & 0.327 & -0.161 & (0.101) & 19 & 0.305\\
China & 0.103 & -0.064 & -0.246$^{*}$ & (0.121) & 30 & 0.118 & -0.059$^{***}$ & (0.016) & 23 & 0.353\\
Colombia & 0.047 & -0.266 & 0.028 & (0.214) & 27 & 0.001 & -0.219$^{***}$ & (0.059) & 24 & 0.483\\
Costa Rica & 0.025 & 0.025 & -0.109$^{*}$ & (0.045) & 8 & 0.219 & -0.063 & (0.065) & 7 & 0.127\\
Croatia & 0.034 & -0.123 & -0.148 & (0.100) & 23 & 0.080 & -0.131$^{***}$ & (0.024) & 21 & 0.682\\
Curaçao & 0.010 & -0.097 & -0.287$^{***}$ & (0.075) & 20 & 0.365 & -0.142$^{***}$ & (0.032) & 20 & 0.561\\
Cyprus & 0.037 & -0.188 & -0.123$^{**}$ & (0.045) & 22 & 0.180 & -0.156$^{***}$ & (0.039) & 24 & 0.617\\
Czechia & 0.008 & -0.105 & -0.318$^{***}$ & (0.048) & 26 & 0.411 & -0.091$^{***}$ & (0.022) & 22 & 0.415\\
Côte d’Ivoire & 0.065 & -0.227 & -0.085 & (0.274) & 32 & 0.003 & -0.285$^{***}$ & (0.063) & 27 & 0.585\\
Denmark & 0.035 & 0.019 & -0.030 & (0.183) & 40 & 0.001 & -0.143$^{***}$ & (0.028) & 38 & 0.413\\
Dominican Republic & 0.057 & -0.191 & -0.210$^{***}$ & (0.055) & 28 & 0.516 & -0.211$^{***}$ & (0.039) & 27 & 0.661\\
Ecuador & 0.053 & -0.164 & -0.137$^{***}$ & (0.028) & 15 & 0.591 & -0.120$^{***}$ & (0.012) & 10 & 0.931\\
Egypt & 0.060 & -0.129 & -0.170 & (0.158) & 22 & 0.170 & -0.174$^{***}$ & (0.045) & 21 & 0.514\\
El Salvador & 0.033 & -0.651 & -0.412$^{***}$ & (0.094) & 7 & 0.864 & -0.439$^{***}$ & (0.054) & 6 & 0.922\\
Estonia & 0.039 & -0.050 & -0.251$^{*}$ & (0.139) & 27 & 0.072 & -0.217$^{*}$ & (0.115) & 23 & 0.334\\
Finland & 0.038 & -0.017 & -0.269$^{***}$ & (0.086) & 36 & 0.125 & -0.130$^{***}$ & (0.043) & 29 & 0.409\\
France & 0.022 & -0.180 & -0.335$^{***}$ & (0.114) & 124 & 0.097 & -0.235$^{***}$ & (0.037) & 88 & 0.476\\
Germany & 0.040 & -0.133 & -0.659$^{***}$ & (0.187) & 119 & 0.215 & -0.161$^{***}$ & (0.022) & 86 & 0.722\\
Greece & 0.018 & -0.100 & -0.124$^{***}$ & (0.028) & 24 & 0.382 & -0.120$^{***}$ & (0.039) & 24 & 0.373\\
Guatemala & 0.037 & -0.165 & -0.212 & (0.137) & 19 & 0.362 & -0.129$^{**}$ & (0.047) & 16 & 0.441\\
Guinea & 0.196 & -0.101 & -0.243$^{**}$ & (0.087) & 19 & 0.300 & -0.282$^{***}$ & (0.049) & 16 & 0.752\\
Honduras & 0.034 & 0.047 & -0.096 & (0.094) & 16 & 0.061 & -0.155$^{**}$ & (0.069) & 15 & 0.372\\
Hong Kong & 0.061 & -0.089 & -0.415$^{***}$ & (0.095) & 41 & 0.282 & -0.099$^{***}$ & (0.026) & 37 & 0.435\\
Hungary & 0.033 & -0.148 & 0.195 & (0.129) & 23 & 0.120 & -0.148$^{***}$ & (0.039) & 20 & 0.408\\
Iceland & 0.037 & -0.143 & -0.069$^{**}$ & (0.027) & 19 & 0.136 & -0.094$^{***}$ & (0.027) & 17 & 0.598\\
India & 0.074 & 0.057 & -0.046 & (0.054) & 23 & 0.014 & -0.072$^{***}$ & (0.011) & 15 & 0.624\\
Indonesia & 0.062 & -0.150 & -0.118 & (0.056) & 5 & 0.492 & -0.112$^{*}$ & (0.037) & 5 & 0.658\\
Iran & 0.029 & -0.088 & 0.026 & (0.052) & 20 & 0.009 & -0.045 & (0.033) & 24 & 0.105\\
Ireland & 0.057 & -0.038 & -0.150$^{**}$ & (0.060) & 25 & 0.287 & -0.121$^{***}$ & (0.035) & 22 & 0.314\\
Israel & 0.054 & -0.198 & -0.251 & (0.255) & 25 & 0.038 & -0.146$^{***}$ & (0.023) & 21 & 0.631\\
Italy & 0.025 & -0.193 & -0.480$^{***}$ & (0.134) & 35 & 0.234 & -0.165$^{***}$ & (0.044) & 19 & 0.534\\
Japan & 0.025 & -0.109 & -0.064 & (0.057) & 28 & 0.045 & -0.058 & (0.042) & 26 & 0.152\\
Kazakhstan & 0.022 & -0.152 & -0.090$^{***}$ & (0.028) & 29 & 0.175 & -0.122$^{***}$ & (0.040) & 28 & 0.543\\
Kuwait & 0.044 & 0.142 & -0.090 & (0.071) & 18 & 0.101 & -0.081$^{**}$ & (0.031) & 17 & 0.216\\
Kyrgyzstan & 0.024 & -0.313 & -0.185$^{***}$ & (0.062) & 31 & 0.253 & -0.145$^{***}$ & (0.027) & 30 & 0.524\\
Latvia & 0.044 & -0.232 & -0.350$^{**}$ & (0.132) & 26 & 0.134 & -0.267$^{***}$ & (0.088) & 26 & 0.441\\
Lithuania & 0.048 & -0.183 & -0.275$^{***}$ & (0.033) & 25 & 0.433 & -0.334$^{***}$ & (0.022) & 20 & 0.837\\
Luxembourg & 0.047 & -0.023 & -0.087 & (0.065) & 24 & 0.041 & -0.061$^{**}$ & (0.024) & 22 & 0.333\\
Malaysia & 0.044 & -0.079 & -0.059 & (0.058) & 16 & 0.044 & -0.107$^{***}$ & (0.014) & 11 & 0.561\\
Malta & 0.043 & -0.009 & 0.090 & (0.079) & 24 & 0.036 & -0.033 & (0.038) & 22 & 0.032\\
Mauritius & 0.029 & -0.068 & -0.060 & (0.033) & 8 & 0.255 & -0.107$^{***}$ & (0.019) & 8 & 0.645\\
Mexico & 0.029 & -0.055 & -0.156$^{*}$ & (0.091) & 25 & 0.099 & -0.201$^{***}$ & (0.034) & 21 & 0.621\\
Moldova & -0.004 & -0.267 & -0.229$^{*}$ & (0.115) & 31 & 0.269 & -0.121$^{***}$ & (0.029) & 32 & 0.557\\
Mongolia & 0.055 & 0.054 & 0.064 & (0.097) & 15 & 0.041 & -0.046 & (0.039) & 15 & 0.136\\
Morocco & 0.054 & -0.220 & -0.214$^{***}$ & (0.062) & 22 & 0.262 & -0.128$^{***}$ & (0.034) & 18 & 0.572\\
Netherlands & 0.026 & -0.099 & -0.067 & (0.120) & 38 & 0.013 & -0.077$^{***}$ & (0.024) & 28 & 0.305\\
New Zealand & 0.044 & -0.131 & -0.063 & (0.066) & 32 & 0.027 & -0.085$^{***}$ & (0.027) & 33 & 0.152\\
Nicaragua & 0.035 & -0.115 & -0.146$^{**}$ & (0.050) & 14 & 0.624 & -0.178$^{***}$ & (0.044) & 14 & 0.663\\
Niger & 0.067 & -0.208 & -0.200 & (0.133) & 27 & 0.099 & -0.188$^{***}$ & (0.040) & 23 & 0.682\\
Norway & 0.046 & -0.037 & -0.232 & (0.201) & 42 & 0.049 & -0.129$^{***}$ & (0.020) & 38 & 0.569\\
Peru & 0.048 & -0.175 & -0.169$^{***}$ & (0.041) & 15 & 0.657 & -0.142$^{***}$ & (0.037) & 12 & 0.731\\
Philippines & 0.050 & -0.235 & -0.220$^{*}$ & (0.119) & 28 & 0.236 & -0.234$^{***}$ & (0.039) & 22 & 0.587\\
Poland & 0.063 & -0.142 & -0.729$^{***}$ & (0.159) & 30 & 0.430 & -0.116$^{***}$ & (0.037) & 21 & 0.561\\
Portugal & 0.030 & -0.149 & -0.305$^{***}$ & (0.072) & 22 & 0.230 & -0.129$^{***}$ & (0.022) & 19 & 0.631\\
Qatar & 0.085 & 0.030 & 0.064$^{**}$ & (0.025) & 20 & 0.204 & -0.005 & (0.050) & 19 & 0.001\\
Romania & 0.011 & -0.071 & -0.313$^{***}$ & (0.098) & 30 & 0.421 & -0.138$^{***}$ & (0.043) & 27 & 0.606\\
Russia & 0.045 & -0.170 & -0.147$^{*}$ & (0.082) & 22 & 0.155 & -0.104$^{***}$ & (0.034) & 24 & 0.409\\
Saudi Arabia & 0.068 & -0.102 & -0.135$^{***}$ & (0.037) & 17 & 0.299 & -0.053 & (0.041) & 14 & 0.128\\
Senegal & 0.051 & -0.116 & -0.119 & (0.063) & 8 & 0.327 & -0.143$^{*}$ & (0.048) & 5 & 0.484\\
Serbia & 0.052 & -0.327 & -0.473$^{**}$ & (0.169) & 25 & 0.221 & -0.307$^{***}$ & (0.062) & 20 & 0.594\\
Slovakia & 0.028 & -0.135 & -0.276$^{***}$ & (0.059) & 24 & 0.504 & -0.109$^{***}$ & (0.025) & 26 & 0.433\\
Slovenia & 0.033 & -0.087 & -0.083 & (0.077) & 26 & 0.029 & -0.150$^{***}$ & (0.037) & 18 & 0.580\\
South Africa & 0.024 & -0.098 & -0.210$^{***}$ & (0.063) & 18 & 0.327 & -0.148$^{***}$ & (0.035) & 23 & 0.444\\
South Korea & 0.080 & -0.153 & -0.285$^{**}$ & (0.119) & 51 & 0.245 & -0.090$^{**}$ & (0.035) & 41 & 0.420\\
Spain & 0.032 & -0.213 & -0.210$^{**}$ & (0.074) & 22 & 0.199 & -0.091$^{**}$ & (0.037) & 21 & 0.232\\
Sweden & 0.041 & -0.055 & 0.086 & (0.133) & 64 & 0.006 & -0.194$^{***}$ & (0.032) & 56 & 0.475\\
Switzerland & 0.032 & -0.078 & 0.056 & (0.154) & 24 & 0.005 & -0.084$^{**}$ & (0.032) & 22 & 0.309\\
Trinidad \& Tobago & 0.031 & -0.140 & -0.329$^{***}$ & (0.048) & 22 & 0.661 & -0.143$^{***}$ & (0.021) & 20 & 0.630\\
Tunisia & 0.041 & -0.117 & -0.249$^{**}$ & (0.097) & 27 & 0.211 & -0.182$^{***}$ & (0.029) & 25 & 0.573\\
Turkey & 0.068 & -0.329 & -0.094 & (0.124) & 13 & 0.030 & -0.200$^{**}$ & (0.067) & 11 & 0.504\\
USA & 0.037 & -0.175 & -0.131$^{**}$ & (0.056) & 142 & 0.032 & -0.125$^{***}$ & (0.031) & 89 & 0.281\\
Ukraine & -0.017 & -0.173 & -0.179$^{***}$ & (0.043) & 30 & 0.265 & -0.125$^{***}$ & (0.029) & 28 & 0.746\\
United Kingdom & 0.021 & -0.107 & -0.086 & (0.155) & 142 & 0.003 & -0.172$^{***}$ & (0.054) & 97 & 0.249\\
Uruguay & 0.005 & 0.087 & -0.138 & (0.128) & 6 & 0.254 & -0.116 & (0.062) & 5 & 0.340\\
Uzbekistan & 0.052 & -0.093 & -0.019 & (0.126) & 12 & 0.002 & -0.149$^{***}$ & (0.020) & 9 & 0.859\\
Vanuatu & 0.048 & -0.034 & 0.025 & (0.105) & 19 & 0.003 & -0.062$^{**}$ & (0.021) & 15 & 0.369\\
Venezuela & -0.023 & -0.001 & 0.026 & (0.028) & 22 & 0.034 & -0.058$^{*}$ & (0.030) & 22 & 0.267\\
ALL COUNTRIES\textsuperscript{a} & 0.033 & -0.152 & -0.121$^{**}$ & (0.058) & 2744 & 0.009 & -0.152$^{***}$ & (0.020) & 2347 & 0.304\\
\bottomrule
\end{longtable}
\begin{flushleft}
\footnotesize \textsuperscript{a} GDP-weighted.\\
\footnotesize \emph{Note:} Three, two and one asterisks denote $p<0.01$, $p<0.05$ and $p<0.10$, respectively.
\end{flushleft}
\endgroup

\clearpage
\begingroup
\fontsize{9}{9.8}\selectfont
\renewcommand{\arraystretch}{1.04}
\setlength{\tabcolsep}{1.5pt}
\setlength{\LTleft}{0pt}
\setlength{\LTright}{0pt}
\setlength{\LTcapwidth}{\linewidth}
\begin{longtable}{lrrrrrrrrccc}
\caption{$g(K)$, $f(K)$, $r(K)$, $\Delta f(K)$, $\Delta r(K)$, $\theta_r$, $\phi_r$ and associated coefficient of regression averaged over all years reported for each country. Screen = 0.01. $H_0$ per thrift theory: $\theta_r = 1$, and coefficient of regression of $-\Delta f(K)$ on $\Delta g(K) = 1$.}\label{tab:free_growth_results}\\
\toprule
Country & $g(K)$ & $f(K)$ & $r(K)$ & $\Delta f(K)$ & $\Delta r(K)$ & $\theta_r$ & $\phi_r$ & Coef. & S.E. & Obs. & R$^2$\\
\midrule
\endfirsthead
\caption*{\textbf{\tablename~\thetable}\ (cont.)}\\
\toprule
Country & $g(K)$ & $f(K)$ & $r(K)$ & $\Delta f(K)$ & $\Delta r(K)$ & $\theta_r$ & $\phi_r$ & Coef. & S.E. & Obs. & R$^2$\\
\midrule
\endhead
Armenia & 0.027 & 0.149 & 0.177 & 0.005 & 0.021 & -0.069 & 1.069 & -0.068$^{***}$ & (0.023) & 31 & 0.266\\
Aruba & 0.024 & 0.081 & 0.105 & 0.000 & 0.006 & -0.055 & 1.055 & -0.056$^{***}$ & (0.009) & 23 & 0.710\\
Australia & 0.049 & 0.044 & 0.093 & -0.000 & -0.001 & -0.001 & 1.001 & -0.024$^{***}$ & (0.007) & 50 & 0.175\\
Austria & 0.027 & 0.051 & 0.078 & -0.000 & -0.003 & -0.035 & 1.035 & -0.028$^{**}$ & (0.010) & 21 & 0.236\\
Azerbaijan & 0.052 & 0.109 & 0.161 & 0.002 & 0.010 & 0.031 & 0.969 & -0.029 & (0.052) & 28 & 0.029\\
Bahrain & 0.034 & 0.074 & 0.108 & 0.002 & 0.024 & -0.046 & 1.046 & -0.033$^{***}$ & (0.009) & 13 & 0.443\\
Belarus & 0.019 & 0.064 & 0.083 & 0.000 & 0.004 & 0.003 & 0.997 & -0.013 & (0.016) & 29 & 0.042\\
Belgium & 0.031 & 0.038 & 0.069 & -0.000 & -0.000 & -0.014 & 1.014 & -0.005 & (0.020) & 29 & 0.004\\
Bolivia & 0.054 & 0.137 & 0.191 & -0.002 & -0.001 & -0.029 & 1.029 & -0.096$^{***}$ & (0.023) & 23 & 0.430\\
Botswana & 0.048 & 0.074 & 0.122 & 0.000 & -0.005 & -0.020 & 1.020 & -0.083$^{***}$ & (0.026) & 30 & 0.384\\
Brazil & 0.025 & 0.100 & 0.124 & -0.001 & -0.000 & -0.051 & 1.051 & -0.042$^{*}$ & (0.024) & 18 & 0.226\\
British Virgin Islands & 0.124 & 0.017 & 0.141 & -0.002 & 0.022 & 0.009 & 0.991 & -0.001 & (0.005) & 23 & 0.005\\
Bulgaria & 0.065 & 0.181 & 0.247 & -0.018 & 0.007 & -0.100 & 1.100 & -0.007 & (0.076) & 19 & 0.002\\
Burkina Faso & 0.055 & 0.095 & 0.151 & -0.000 & -0.002 & -0.066 & 1.066 & -0.038$^{*}$ & (0.019) & 21 & 0.172\\
Cameroon & 0.045 & 0.126 & 0.171 & 0.001 & 0.008 & -0.134 & 1.134 & -0.070$^{***}$ & (0.021) & 26 & 0.511\\
Canada & 0.049 & 0.066 & 0.115 & -0.001 & -0.002 & 0.035 & 0.965 & -0.009 & (0.028) & 43 & 0.003\\
Cape Verde & 0.027 & 0.106 & 0.134 & 0.000 & 0.009 & -0.069 & 1.069 & -0.088$^{***}$ & (0.005) & 11 & 0.930\\
Chile & 0.037 & 0.099 & 0.136 & 0.001 & -0.003 & -0.111 & 1.111 & -0.091$^{**}$ & (0.039) & 19 & 0.334\\
China & 0.103 & 0.011 & 0.114 & -0.001 & -0.009 & -0.004 & 1.004 & -0.008 & (0.009) & 23 & 0.038\\
Colombia & 0.047 & 0.155 & 0.202 & 0.001 & 0.008 & -0.127 & 1.127 & -0.117$^{***}$ & (0.031) & 24 & 0.441\\
Costa Rica & 0.025 & 0.080 & 0.105 & -0.001 & 0.000 & -0.024 & 1.024 & -0.004 & (0.045) & 7 & 0.001\\
Croatia & 0.034 & 0.069 & 0.103 & -0.002 & 0.002 & -0.060 & 1.060 & -0.038 & (0.025) & 21 & 0.178\\
Curaçao & 0.010 & 0.047 & 0.057 & -0.001 & -0.001 & 0.020 & 0.980 & -0.005 & (0.017) & 20 & 0.005\\
Cyprus & 0.037 & 0.097 & 0.134 & 0.000 & 0.001 & -0.079 & 1.079 & -0.081$^{**}$ & (0.036) & 24 & 0.400\\
Czechia & 0.008 & 0.045 & 0.053 & 0.001 & 0.013 & 0.010 & 0.990 & -0.014 & (0.011) & 22 & 0.051\\
Côte d’Ivoire & 0.065 & 0.186 & 0.251 & -0.000 & 0.004 & -0.077 & 1.077 & -0.109$^{***}$ & (0.033) & 27 & 0.410\\
Denmark & 0.035 & 0.062 & 0.097 & -0.001 & -0.001 & -0.002 & 1.002 & -0.051$^{***}$ & (0.008) & 38 & 0.465\\
Dominican Republic & 0.057 & 0.135 & 0.192 & 0.001 & 0.003 & -0.032 & 1.032 & -0.071$^{***}$ & (0.025) & 27 & 0.339\\
Ecuador & 0.053 & 0.050 & 0.103 & 0.001 & 0.007 & -0.065 & 1.065 & -0.047$^{***}$ & (0.006) & 10 & 0.819\\
Egypt & 0.060 & 0.240 & 0.300 & -0.000 & -0.001 & -0.047 & 1.047 & -0.098$^{**}$ & (0.038) & 21 & 0.321\\
El Salvador & 0.033 & 0.224 & 0.258 & 0.006 & 0.018 & -0.613 & 1.613 & -0.288$^{***}$ & (0.033) & 6 & 0.860\\
Estonia & 0.039 & 0.074 & 0.113 & 0.000 & 0.002 & 0.045 & 0.955 & -0.032 & (0.041) & 23 & 0.055\\
Finland & 0.038 & 0.065 & 0.103 & -0.000 & -0.004 & -0.011 & 1.011 & -0.022 & (0.019) & 29 & 0.108\\
France & 0.023 & 0.068 & 0.091 & 0.000 & 0.000 & -0.043 & 1.043 & -0.079$^{***}$ & (0.021) & 83 & 0.215\\
Germany & 0.037 & 0.049 & 0.086 & 0.001 & -0.009 & -0.030 & 1.030 & -0.015$^{**}$ & (0.005) & 34 & 0.085\\
Greece & 0.018 & 0.105 & 0.123 & 0.000 & 0.004 & -0.022 & 1.022 & -0.024 & (0.015) & 24 & 0.099\\
Guatemala & 0.037 & 0.145 & 0.182 & 0.002 & 0.006 & -0.040 & 1.040 & -0.066 & (0.046) & 16 & 0.178\\
Guinea & 0.196 & 0.154 & 0.349 & -0.003 & 0.003 & -0.082 & 1.082 & -0.056$^{**}$ & (0.025) & 16 & 0.246\\
Honduras & 0.034 & 0.095 & 0.129 & 0.003 & 0.011 & 0.016 & 0.984 & -0.079$^{*}$ & (0.039) & 15 & 0.313\\
Hong Kong & 0.061 & 0.041 & 0.103 & -0.001 & -0.008 & -0.006 & 1.006 & -0.021 & (0.012) & 37 & 0.064\\
Hungary & 0.033 & 0.067 & 0.100 & 0.000 & 0.006 & -0.013 & 1.013 & -0.040$^{**}$ & (0.018) & 20 & 0.164\\
Iceland & 0.037 & 0.050 & 0.087 & 0.000 & 0.002 & -0.036 & 1.036 & -0.016 & (0.018) & 17 & 0.057\\
India & 0.074 & 0.053 & 0.128 & 0.002 & 0.014 & 0.056 & 0.944 & -0.020$^{***}$ & (0.005) & 15 & 0.226\\
Indonesia & 0.062 & 0.044 & 0.106 & 0.001 & 0.016 & -0.031 & 1.031 & -0.023 & (0.016) & 5 & 0.337\\
Iran & 0.029 & 0.042 & 0.071 & 0.000 & 0.003 & -0.061 & 1.061 & -0.002 & (0.018) & 24 & 0.001\\
Ireland & 0.057 & 0.049 & 0.106 & -0.000 & -0.007 & -0.022 & 1.022 & -0.031$^{**}$ & (0.013) & 22 & 0.145\\
Israel & 0.054 & 0.089 & 0.143 & -0.002 & -0.003 & -0.069 & 1.069 & -0.035 & (0.028) & 21 & 0.145\\
Italy & 0.025 & 0.067 & 0.093 & -0.001 & -0.006 & -0.046 & 1.046 & -0.048$^{***}$ & (0.015) & 19 & 0.403\\
Japan & 0.025 & 0.033 & 0.057 & 0.000 & -0.000 & -0.020 & 1.020 & -0.021$^{*}$ & (0.011) & 26 & 0.192\\
Kazakhstan & 0.022 & 0.068 & 0.090 & 0.001 & 0.015 & -0.027 & 1.027 & -0.019 & (0.026) & 28 & 0.035\\
Kuwait & 0.044 & 0.035 & 0.079 & 0.000 & -0.010 & 0.175 & 0.825 & -0.045 & (0.035) & 17 & 0.103\\
Kyrgyzstan & 0.024 & 0.132 & 0.155 & 0.004 & 0.016 & -0.208 & 1.208 & -0.075$^{***}$ & (0.021) & 30 & 0.274\\
Latvia & 0.044 & 0.167 & 0.210 & -0.003 & -0.004 & -0.049 & 1.049 & -0.076 & (0.046) & 26 & 0.105\\
Lithuania & 0.048 & 0.120 & 0.169 & -0.003 & -0.012 & -0.028 & 1.028 & -0.130$^{***}$ & (0.033) & 20 & 0.461\\
Luxembourg & 0.047 & 0.039 & 0.086 & -0.002 & -0.003 & 0.037 & 0.963 & -0.022 & (0.014) & 22 & 0.091\\
Malaysia & 0.044 & 0.062 & 0.106 & 0.001 & -0.000 & 0.015 & 0.985 & -0.029 & (0.030) & 11 & 0.108\\
Malta & 0.043 & 0.053 & 0.095 & -0.001 & -0.002 & 0.064 & 0.936 & 0.007 & (0.023) & 22 & 0.002\\
Mauritius & 0.029 & 0.106 & 0.135 & -0.002 & -0.041 & -0.017 & 1.017 & -0.057$^{***}$ & (0.011) & 8 & 0.584\\
Mexico & 0.029 & 0.113 & 0.142 & -0.000 & 0.003 & -0.005 & 1.005 & -0.092$^{***}$ & (0.015) & 21 & 0.472\\
Moldova & -0.004 & 0.110 & 0.107 & 0.003 & -0.001 & -0.100 & 1.100 & -0.039$^{**}$ & (0.015) & 32 & 0.222\\
Mongolia & 0.055 & 0.068 & 0.123 & 0.001 & -0.003 & 0.040 & 0.960 & 0.002 & (0.017) & 15 & 0.001\\
Morocco & 0.054 & 0.083 & 0.137 & 0.001 & -0.005 & -0.067 & 1.067 & -0.049$^{***}$ & (0.012) & 18 & 0.523\\
Netherlands & 0.026 & 0.038 & 0.064 & -0.000 & -0.000 & -0.019 & 1.019 & 0.011 & (0.015) & 28 & 0.023\\
New Zealand & 0.044 & 0.017 & 0.061 & 0.000 & -0.000 & -0.064 & 1.064 & -0.019 & (0.033) & 33 & 0.006\\
Nicaragua & 0.035 & 0.095 & 0.130 & -0.001 & 0.002 & 0.034 & 0.966 & -0.072$^{**}$ & (0.027) & 14 & 0.439\\
Niger & 0.067 & 0.120 & 0.187 & -0.001 & -0.000 & -0.068 & 1.068 & -0.044$^{**}$ & (0.017) & 23 & 0.284\\
Norway & 0.046 & 0.061 & 0.107 & -0.001 & -0.007 & 0.019 & 0.981 & -0.037$^{***}$ & (0.008) & 38 & 0.313\\
Peru & 0.048 & 0.084 & 0.132 & 0.002 & 0.013 & -0.064 & 1.064 & -0.054$^{***}$ & (0.015) & 12 & 0.615\\
Philippines & 0.050 & 0.103 & 0.153 & 0.002 & 0.007 & -0.057 & 1.057 & -0.050 & (0.038) & 22 & 0.075\\
Poland & 0.063 & 0.085 & 0.148 & -0.002 & -0.004 & 0.012 & 0.988 & 0.000 & (0.030) & 21 & 0.000\\
Portugal & 0.030 & 0.059 & 0.089 & -0.001 & -0.001 & -0.044 & 1.044 & -0.050$^{***}$ & (0.014) & 19 & 0.434\\
Qatar & 0.085 & 0.031 & 0.116 & -0.000 & -0.005 & 0.007 & 0.993 & -0.003 & (0.012) & 19 & 0.004\\
Romania & 0.011 & 0.080 & 0.090 & 0.003 & 0.006 & -0.074 & 1.074 & -0.031 & (0.024) & 27 & 0.136\\
Russia & 0.045 & 0.055 & 0.100 & 0.001 & -0.005 & -0.060 & 1.060 & -0.005 & (0.016) & 24 & 0.007\\
Saudi Arabia & 0.068 & 0.057 & 0.125 & 0.001 & 0.003 & 0.004 & 0.996 & -0.019 & (0.028) & 14 & 0.040\\
Senegal & 0.051 & 0.148 & 0.199 & -0.002 & -0.027 & -0.054 & 1.054 & -0.040 & (0.020) & 5 & 0.253\\
Serbia & 0.052 & 0.196 & 0.248 & -0.006 & -0.007 & -0.105 & 1.105 & -0.149$^{***}$ & (0.037) & 20 & 0.512\\
Slovakia & 0.028 & 0.081 & 0.109 & 0.001 & 0.008 & -0.062 & 1.062 & -0.031 & (0.023) & 26 & 0.117\\
Slovenia & 0.033 & 0.047 & 0.080 & 0.000 & 0.005 & -0.009 & 1.009 & -0.029 & (0.037) & 18 & 0.064\\
South Africa & 0.024 & 0.100 & 0.124 & -0.000 & -0.000 & -0.039 & 1.039 & -0.057$^{**}$ & (0.023) & 23 & 0.231\\
South Korea & 0.080 & 0.034 & 0.114 & -0.001 & -0.003 & -0.064 & 1.064 & -0.005 & (0.024) & 41 & 0.004\\
Spain & 0.032 & 0.040 & 0.072 & -0.000 & -0.001 & -0.048 & 1.048 & -0.022 & (0.017) & 21 & 0.095\\
Sweden & 0.041 & 0.068 & 0.109 & -0.000 & 0.001 & -0.005 & 1.005 & -0.034$^{**}$ & (0.016) & 56 & 0.083\\
Switzerland & 0.032 & 0.012 & 0.044 & -0.000 & 0.001 & -0.027 & 1.027 & -0.041$^{***}$ & (0.008) & 22 & 0.457\\
Trinidad \& Tobago & 0.031 & 0.074 & 0.105 & -0.001 & -0.010 & -0.050 & 1.050 & -0.050$^{***}$ & (0.009) & 20 & 0.578\\
Tunisia & 0.041 & 0.114 & 0.154 & 0.000 & 0.007 & -0.065 & 1.065 & -0.066$^{**}$ & (0.026) & 25 & 0.284\\
Turkey & 0.068 & 0.137 & 0.205 & -0.005 & -0.015 & -0.176 & 1.176 & -0.122$^{*}$ & (0.063) & 11 & 0.260\\
USA & 0.034 & 0.055 & 0.089 & 0.000 & -0.001 & -0.053 & 1.053 & -0.039$^{***}$ & (0.011) & 78 & 0.218\\
Ukraine & -0.017 & 0.062 & 0.045 & 0.001 & -0.008 & -0.034 & 1.034 & -0.043$^{***}$ & (0.013) & 28 & 0.511\\
United Kingdom & 0.021 & 0.070 & 0.091 & 0.002 & 0.001 & 0.056 & 0.944 & -0.003 & (0.030) & 97 & 0.000\\
Uruguay & 0.005 & 0.083 & 0.088 & 0.001 & 0.006 & 0.062 & 0.938 & -0.035 & (0.027) & 5 & 0.170\\
Uzbekistan & 0.052 & 0.100 & 0.152 & 0.006 & -0.001 & -0.015 & 1.015 & -0.026 & (0.039) & 9 & 0.075\\
Vanuatu & 0.048 & 0.026 & 0.074 & -0.000 & 0.006 & 0.025 & 0.975 & 0.023 & (0.023) & 15 & 0.083\\
Venezuela & -0.023 & 0.061 & 0.038 & 0.001 & 0.003 & 0.025 & 0.975 & -0.016 & (0.028) & 22 & 0.030\\
ALL COUNTRIES\textsuperscript{a} & 0.032 & 0.053 & 0.085 & 0.001 & -0.001 & -0.009 & 1.009 & -0.033$^{***}$ & (0.011) & 2279 & 0.050\\
\bottomrule
\end{longtable}
\begin{flushleft}
\footnotesize \textsuperscript{a} GDP-weighted.\\
\footnotesize \emph{Note:} Three, two and one asterisks denote $p<0.01$, $p<0.05$ and $p<0.10$, respectively.
\end{flushleft}
\endgroup

\clearpage

\FloatBarrier

\begin{figure}[p]
    \centering
    \includegraphics[width=\textwidth]{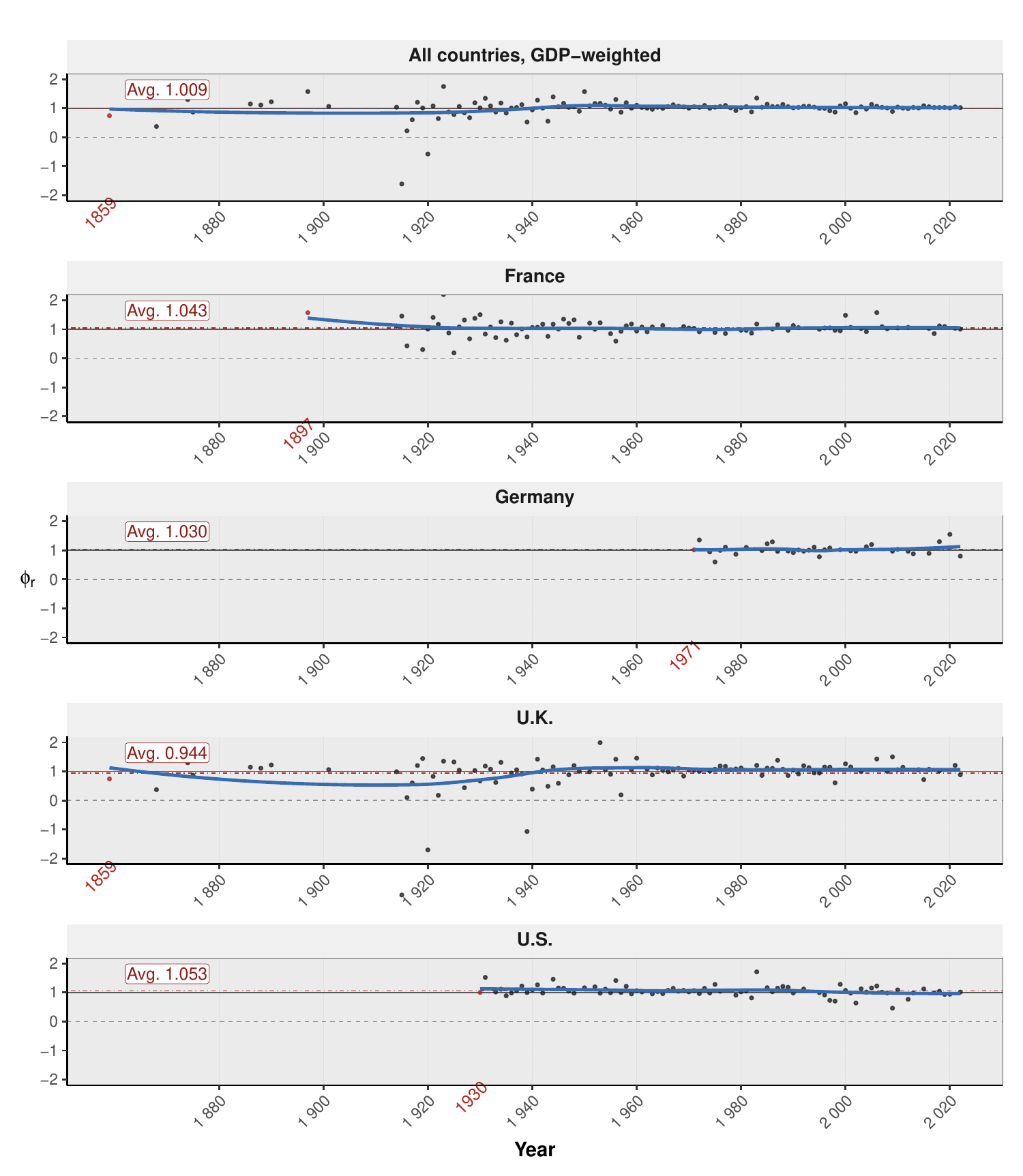}
    \caption{Free growth index ($\phi_r$) for all years in countries shown. Screen = 0.01. Dashed line shows the thrift theory prediction $\phi_r = 0$. Blue curves show LOESS best fit; red dot-dash lines show panel averages.}
    \label{fig-free-growth-indicators}
\end{figure}

\FloatBarrier

\section{Test findings overall}\label{test-findings}

Findings from this method appear in Table~\ref{tab:free_growth_results} and Fig.~\ref{fig-free-growth-indicators}. The composite free growth index \(\phi_r\), averaged over all years and all countries, GDP-weighted, was found at 1.009. Overall thrift index \(\theta_r\), its complement to unity, shows as -0.009. For consistency with Table~\ref{tab:interpretive_results}, we show regressions of \(- \Delta f (K)\) on \(\Delta g (K)\) rather than of \(\Delta r(K)\) on \(\Delta g(K)\). The GDP-weighted pooled regression coefficient was found at -0.033.

Sections~\ref{test-results} and \ref{sep12-consumption-restraint} did not yet
define a free growth index, but rather tested thrift theory by comparing
capital growth to net saving and consumption restraint respectively.
We found $\theta=0.016$ and $\theta_c=-0.152$.
Regressions of $s^*$ on $g(K)$, of $\Delta s^*$ on $\Delta g(K)$,
of $-c^*$ on $g(K)$ and of $-\Delta c^*$ on $\Delta g(K)$ were found at
$0.141$, $0.018$, $-0.121$ and $-0.152$, respectively.
We propose that all these findings together refute thrift theory, and
establish free growth theory as a general principle until more is known.

\section{Comparison to traditional measurements of $r(K)$}
\label{oct-comparison-traditional-returns}

Rate of return, within risk tolerance, is the maximand of investment.
Meanwhile wars have been fought about its justifiability.
On both grounds, it has been one of the most studied of economic variables.
It has been measured at market for particular classes of assets, then
cap-weighted where practical to give return overall.

That market-based approach offers the advantage of direct observation
of return in those particular asset classes, as listed in
Table~\ref{tab:oct7-jorda-comparison} for example.
Consumption, pay and market-value capital, the determinants from which
we reason in Eqs.~\eqref{eq:total-return} and
\eqref{eq:oct4-cash-flow-consumption}, cannot be measured as closely.
Our method offers more comprehensive coverage, however, as it reports
also for normally untraded wealth such as infrastructure, and it
cap-weights automatically. The two approaches are complementary
rather than competitive.

The market-based study we choose for comparison is Jord\`a, Knoll, Kuvshinov, Schularick
and Taylor (JKKST), published
in \citeyear{jorda2019} but updated through 2020. JKKST
reports for 16 countries over the period
1896 (France and UK alone) through 2020.
Their findings are compared to ours in Tables~\ref{tab:oct7-jorda-comparison} and \ref{tab:oct7-jorda-wealth}.

% PDF page 16 -- "insert pt 2 in p14"; intended manuscript Table 4.
% Move our r(K) column after the JST asset-return columns, following
% the circled column and arrow to the right edge of the table.
\begin{table}[pos=H]
\centering
\small
\caption{Mean annual real JKKST returns compared with our $r(K)$.}
\label{tab:oct7-jorda-comparison}
\begin{tabular}{lrrrrrrr}
\toprule
Country & Years & Obs. & Equities & Housing & Bills & Bonds & Our $r(K)$\\
\midrule
Australia & 1960--2020 & 61 & 7.47 & 6.82 & 1.84 & 3.37 & 9.17\\
Belgium & 1986--2020 & 35 & 8.98 & 8.33 & 1.43 & 5.66 & 6.99\\
Denmark & 1982--2020 & 39 & 12.76 & 6.23 & 2.36 & 7.02 & 10.01\\
Finland & 1981--2020 & 40 & 16.19 & 8.93 & 2.19 & 5.38 & 10.46\\
France & 1896--2020 & 118 & 3.56 & 7.23 & -1.19 & 0.55 & 9.62\\
Germany & 1970--2020 & 51 & 7.38 & 4.75 & 1.44 & 3.62 & 8.64\\
Italy & 1981--2020 & 40 & 7.76 & 4.16 & 2.20 & 6.52 & 9.40\\
Japan & 1981--2020 & 40 & 5.90 & 3.89 & 1.21 & 4.68 & 5.64\\
Netherlands & 1981--2020 & 40 & 11.49 & 7.28 & 1.58 & 5.52 & 6.48\\
Norway & 1979--2020 & 42 & 12.76 & 9.22 & 1.15 & 4.51 & 11.61\\
Portugal & 1996--2020 & 25 & 5.17 & 7.01 & 0.23 & 7.01 & 8.88\\
Spain & 1996--2020 & 24 & 9.40 & 4.38 & 0.03 & 6.27 & 7.43\\
Sweden & 1951--2020 & 70 & 11.19 & 8.77 & 0.62 & 2.79 & 10.98\\
Switzerland & 1991--2020 & 30 & 10.25 & 5.67 & 0.13 & 4.21 & 4.52\\
USA & 1929--2020 & 92 & 8.03 & 5.82 & 1.10 & 2.75 & 8.87\\
United Kingdom & 1896--2020 & 118 & 6.71 & 5.39 & 1.04 & 1.94 & 10.35\\
\bottomrule
\end{tabular}
\end{table}

% PDF page 17 -- "insert pt 3 in p14"; intended manuscript Table 5.
% Move JST wealth before our r(K), following the arrow above the columns.
% Remove the struck "p.p." heading qualifier and "in percentage points".
\begin{table}[pos=H]
\centering
\small
\caption{Comparison with JKKST composite wealth returns on matched country-years (percent per year).}
\label{tab:oct7-jorda-wealth}
\begin{tabular}{lrrrrr}
\toprule
Country & Years & Obs. & JKKST wealth & Our $r(K)$ & Gap\\
\midrule
Australia & 1960--2020 & 61 & 6.84 & 9.17 & 2.33\\
Belgium & 1986--2020 & 35 & 6.97 & 6.99 & 0.02\\
Denmark & 1982--2020 & 39 & 7.59 & 10.01 & 2.41\\
Finland & 1981--2019 & 39 & 11.03 & 10.47 & -0.56\\
France & 1896--2020 & 118 & 4.72 & 9.62 & 4.89\\
Germany & 1970--2020 & 51 & 4.72 & 8.64 & 3.92\\
Italy & 1981--2020 & 40 & 4.68 & 9.40 & 4.72\\
Japan & 1981--2020 & 40 & 4.21 & 5.64 & 1.43\\
Netherlands & 1981--2020 & 40 & 7.37 & 6.48 & -0.89\\
Norway & 1979--2019 & 41 & 8.89 & 11.50 & 2.61\\
Portugal & 1996--2019 & 24 & 6.26 & 9.01 & 2.75\\
Spain & 1996--2017 & 22 & 4.89 & 7.73 & 2.84\\
Sweden & 1951--2019 & 69 & 8.69 & 10.95 & 2.26\\
Switzerland & 1991--2015 & 25 & 7.17 & 4.42 & -2.75\\
USA & 1929--2020 & 92 & 5.33 & 8.87 & 3.54\\
United Kingdom & 1896--2019 & 117 & 4.74 & 10.39 & 5.65\\
\bottomrule
\end{tabular}
\par\medskip
\begin{minipage}{\linewidth}\footnotesize
In both Table~\ref{tab:oct7-jorda-comparison} and Table~\ref{tab:oct7-jorda-wealth}, all
year ranges may include gaps. All
entries report for the same available
years in each country. Coverage
can differ between tables. Entries are
arithmetic means.
Gap is our $r(K)$ less JKKST composite wealth return.
\end{minipage}
\end{table}

\FloatBarrier

Our findings for $r(K)$ clearly exceed JKKST's composite wealth return
in 12 of these 16 countries. Findings are substantially the same as to
Belgium, where we found $r(K)$ at 6.99\% while JKKST found composite wealth
return at 6.97\%. JKKST found higher measurements than ours for Finland,
Netherlands and Switzerland.

For the United States in 1929--2020, we found $r(K)$ at 8.87\%. 
JKKST found 8.03\% for equities, 5.82\% for housing and 5.33\%
for composite wealth return. For Germany in 1970--2020, the
corresponding figures are 8.64\%, 7.38\%, 4.75\% and 4.72\%.

\section{Inferences for creative destruction}\label{sep12-creative-destruction}

New technology obsoletes old technology. Cars and word processors have largely replaced horse-drawn buggies and typewriters. Joseph \citet{schumpeterCapitalismSocialismDemocracy1942} called this process creative destruction. Data and arguments presented here suggest that payment for the new tech (as cars and word processors) came entirely from depreciation of the old tech (as buggies and typewriters) at the collective scale.

Present value, meanwhile, is time-discounted value of expected cash flows. Depreciation is expected loss of value to asset age, taken into account when projecting future cash flows, and is recovered within positive cash flow available for spending. Unexpected loss of capital value is deadweight loss, not depreciation, and thus is unrecovered in cash flow.

Sellers tend to find more buyers at higher prices when buyers have more means of investment, including depreciation investment, and thus tend to avoid causing unexpected loss of value to potential buyers in the form of premature obsolescence. This basic tenet of supply and demand gives sellers a powerful money motive to phase in the new tech at the depreciation rate of the old tech. Replacing the old tech as it wears out is good business, and reducing its value before, by premature obsolescence, is bad business.

These considerations suggest that the creative part of creative destruction is real; but that the destruction part might be misinterpreted. It is possible for obsoleted industries to go out of business smoothly without loss. We argue that both buyers and sellers have motives to favor this result. If evidence from rust belts suggests a less graceful exit, we propose other causes than market motives.

Government is best positioned to foresee and manage negative externalities, most free to represent all sides impartially, and most clearly charged to do so. We suggest that governments might make this job easier by considering that policies to favor investment over consumption might lead to capital glut, and that capital glut might have something to do with rust belts.

\section{Inferences for general welfare}\label{sep12-general-welfare}

Reasonable people can disagree as to what role government should take in equalizing incomes. But it seems to us that governments of all philosophies and motives are likeliest to stay in power, and likeliest to be proud of themselves, when bellies are full and taxes are light.

Data presented here suggest that net saving, at the collective scale, is sacrifice of consumption with nothing in return but capital glut. Suppose that the flow were converted into more consumption and tax relief. More consumption supply would tend to make consumption cheaper, and so to leave less need of government subsidies. Lighter taxes should make consumption more affordable still, and all at no cost to capital growth.

\section{Free growth as a flow}\label{sep12-free-growth-flow}

The marginalists taught that market forces should tend to equalize product price to producer cost, when producer cost includes interest\footnote{On all capital applied, even if owned free and clear.} as the cost of time. This reasoning would explain thrift theory. We accept both in the absence of innovation.

Innovation brings an exception. Innovators find ways to make more productive uses out of existing skills and plant and materials, and so to realize more product value per dollar invested by the innovator before the market catches up. We call this temporary advantage the innovator's reserve.\footnote{\citet{getty2025sourcescapitalgrowth}.} We use the term free growth to mean this flow to innovators plus any spillover or other net positive externalities\footnote{Net of negative externalities.} brought to third parties by the innovation.\footnote{Ibid.}

\section{Inferences for national accounts}\label{sep12-national-accounts}

Substitution of Eq.~\eqref{eq:oct4-cash-flow-consumption} into Eq.~\eqref{eq:total-return} gives $P=\Delta K+C-\Pi$. This can be rearranged as
\begin{equation}
\Pi+P=\Delta K+C, \quad \text{collectively.\footnotemark}
\label{eq:oct17-24}
\end{equation}
\footnotetext{Note that this needs no adjustment for free growth, which is captured on both sides in \(P\) and \(\Delta K\) respectively.}
% Handwritten insertion above the comma after C: ``3 spaces''.
Macro and national accounts take pay plus gross profit as the overall source of payment, called expenditure, on consumption and gross investment. The sum of pay and net profit is equated to net expenditure, meaning net of depreciation offset, and also to the sum of consumption and net investment. Any of these three measures is called net product, in versions national or domestic.
% PROOFREAD: ``depreciation offset'' is faint; retained as read.

We agree that the dispositions of net expenditure are in consumption and net investment. But we have shown above that expenditure on net investment, at the collective scale, contributes substantially
% PROOFREAD: ``depreciation offset'' is faint; retained as read.
% Source PDF page 23, yellow handwritten sheet; continuation of preceding sentence.
% Circled instruction: ``insert pt 2 on p 16''.
nothing to capital growth, and hence to product in any material sense.

% PROOFREAD: The source refers to Eq. (22) below, though the equal sums
% just derived are in Eq. (24). The handwritten reference is retained.
We suggest that net product be redefined as either of the two equal sums in Eq.~\eqref{eq:oct17-24}. If $\Psi_{\mathrm{net}}$ shows net product, that is, we suggest the redefinition
\begin{equation}
\Psi_{\mathrm{net}}=\Pi+P=\Delta K+C, \quad \text{collectively,}
\label{eq:oct17-25}
\end{equation}
% Handwritten insertion above the comma after C: ``3 sp.''.
% PROOFREAD: The final sentence refers to Eq. (23) in the source; retained,
% though the intended reference may be the redefinition in Eq. (25).
without reference to expenditure or saving or investment. Note that Eq.~\eqref{eq:oct17-25} does not change current meanings, as net investment is intended to give $\Delta K$.

Macro, although not national accounts in themselves, goes further in
equating gross and net product to gross and net output, where these
in principle are defined as gross and net value added. Here we differ.
\citet{getty2025sourcescapitalgrowth}
argued that equating the above sums to value added would overlook elements of
value added to human capital.

\section{Why the whole and not the parts?}

The fallacy of composition means an assumption that what is true of each is
true of all. A classic example is in population age. Each individual ages one
year per year, and yet the average age of all together may rise or fall or
hold the same. The reason is that death rates syphon off older cohorts, while
birth rates replenish youngest ones.

Free growth theory expects that net saving does not contribute to capital
growth at the collective scale, but might do so at smaller scales. The
explanation is something like that for individual and collective aging.
Households dissave in consumption, and firms dissave through dividends and
other distributions to owners. Younger households tend to save to afford children and retirement, while older
households tend to dissave to afford both as they arrive.\footnote{\citet{modigliani1954utility}.}
Firms dependent on old technology may show net dissaving, as with buggies and
typewriters, and firms offering new technology may show net saving in
replacing them.

Free growth theory predicts at the collective scale alone, and allows that
net saving might realize capital growth for individual households and firms
if other households and firms dissave to match.

\section{Summary and discussion}\label{sep12-summary-discussion}

This study is meant to supplement rather than replace our earlier ones. It reasons from data from the WID and PZ websites combined, preferring the former where findings differ. It tests thrift theory in a new way from data for consumption and pay, and suggests a larger role for the cash flow concept in economic analysis. 

We claim no originality for the free growth idea in itself.
\citet{getty2025sourcescapitalgrowth} cites precedents in
John \citet{raeNewPrinciplesPolitical1834}, \citet{millPrinciplesPoliticalEconomy1848},
Robert \citet{solowTECHNICALCHANGEAGGREGATE1957},
Thomas Piketty and Gabriel Zucman \citeyearpar{picketyCapitalIsBack2014} and
Mordecai \citet{kurz2023market}.

The last two studies, like ours, compare $s^*$ to $g(K)$, and find
as we do that $g(K)$ is substantially larger. We take another step
by finding $\Delta s^*/\Delta g(K)$, and by measuring regressions of
$s^*$ on $g(K)$ and of $\Delta s^*$ on $\Delta g(K)$. Our findings
from these tests have persuaded us that free growth accounts for
substantially all of capital growth at the collective scale.

We do not claim originality in this point either. It seems to us
that either Rae or Mill could have meant the same. Conventions of
the time limited their use of math, however, and leave us unsure.

% PDF page 26 -- circled instruction: "insert pt 2 in p17".
We are not aware of any other studies arriving at
Eq.~\eqref{eq:oct4-cash-flow-consumption}, and we doubt that we could
have reached it other than through cash flow analysis in
Eqs.~\eqref{eq:value-added} through \eqref{eq:oct17-23}.
This approach also gave us Eqs.~\eqref{eq:oct17-24} and
\eqref{eq:oct17-25}, allowing adjustment of national accounts
without basic reconstruction. 

\bibliographystyle{cas-model2-names}
\bibliography{cas-refs}

\end{document}